\documentclass[amstex,epsf,amssymb,usenatbib]{mnras}
\usepackage{epsf,graphics,subfig}
\usepackage{amsmath,amssymb,natbib}
\usepackage{lmodern}
\usepackage{cite}

\usepackage{rotating,times,graphicx,latexsym,color,longtable,lscape} 

\def\aj{AJ}                   
\def\apj{ApJ}                 
\def\apjl{ApJ}                
\def\apjs{ApJS}               
\def\aap{A\&A}                
\def\mnras{MNRAS}             
\def\pasj{PASJ}               

\title[Simulations of binary YSO outflows]{Binary outflows from young stars: interaction of co-orbital jet and wind }
    
\author[C.J.R. Lynch, M.D. Smith \& S.C.O. Glover]
{ Chris  J.R. Lynch $^{1}$\thanks{E-mail: captainlockheed@btopenworld.com }, 
{Michael D. Smith $^{1}$\thanks{E-mail: m.d.smith@kent.ac.uk }  \& 
Simon C.O. Glover$^{2}$\thanks{E-mail: glover@uni-heidelberg.de } }\\
$^{1}$Centre for Astrophysics \& Planetary Science, The University of Kent, Canterbury, Kent CT2 7NH, U.K. \\
$^{2}$Universit\"at Heidelberg, Zentrum f\"ur Astronomie, Institut fr theoretische Astrophysik, Albert-Ueberle-Str. 2, 69120 Heidelberg, Germany 
}                                                                                                                                                             
\date{Accepted .....
      Received ..... ;
      in original form .....}
                                                                                                                                                                  
\pagerange{\pageref{firstpage}--\pageref{lastpage}}
\pubyear{2019}

\begin{document}
                                                                                                                                                             
\maketitle
                                                                                                                                                             
\label{firstpage}
                                                                                                                                                             
\begin{abstract}
Jets from young stellar objects provide insight into the workings  of the beating heart at the centre of star forming cores. In some cases, multiple pulsed outflows are detected such as the atomic and molecular jets from a proposed binary system in the T\,Tauri star HH\,30.
We investigate here the development and propagation of  duelling atomic and molecular outflows stemming from the two stars in co-orbit. 
We perform a series of numerical experiments with the {\small ZEUS-MP} code with enhanced cooling and chemistry modules.
 The aim of this work is to identify signatures on scales of order 100\,AU.
The jet sources are off the grid domain and so it is the propagation and interaction from $\sim$ 20\,AU out to 100\,AU simulated here.
 We find that the molecular flow from the orbiting source significantly disturbs  the atomic jet, deflecting and twisting the jet and disrupting the jet knots.
 Regions of high ionisation are generated as the atomic jet rams through the dense molecular outflow. Synthetic images in atomic and molecular lines are presented
 which demonstrate identifying signatures.
  In particular, the structure within the atomic jet is lost and H$\alpha$ may trace the walls of the present CO cavity or where the walls have been recently. 
 These results provide a framework for the interpretation of upcoming 
 high resolution observations.
 
\noindent 
\end{abstract} 
  
\section{Introduction} 
\label{intro}

Protostellar jets offer an observational window into the births of stars. The jets protrude from the 
 cloud which obscures the protostar and remain prominent even when the young star is visible. 
The characteristics of these jets may reveal something of the nature of the originating objects and the
processes that govern their evolution. For example, reflection symmetry and mirror symmetry may differentiate between 
orbital dynamics  and precessional motion such as emphasized in recent studies  \citep{2016MNRAS.460.1829M,2019MNRAS.485.4667H}.
In addition to the role these jets play in extracting angular momentum from infalling 
material in the accretion disc, the emerging outflows may also act as a feedback channel
supplying turbulent energy to the surrounding molecular cloud, which will affect the star
formation efficiency within the cloud  \citep{2007prpl.conf..245A,2017A&A...597A..64D}

Evidence for interacting multiple jet components introduces new challenges. The T\,Tauri star associated with 
HH\,30  \citep{1983ApJ...274L..83M} is a well-known example. Each  jet component can be observed in atomic or 
molecular tracers in the form of distinct knots within a diffuse channel. These components
display behaviour which could be driven by orbiting binary stars. In fact, observations provide evidence that HH\,30 is a 
binary system, surrounded by a circumbinary accretion disc.
Other outflows appear to be driven by multiple sources which may be gravitationally bound within a protostellar core 
\citep{1991ApJ...376..615A,2001A&A...375.1018G}. Twin jets are also found in the L1551\,IRS5 system 
\citep{1998ApJ...499L..75F} and in L\,1157  \citep{2015ApJ...814...43K}. In addition, C-shaped bending where two protostars are present \citep{2009ApJ...699.1584L} could be attributed to an orbiting jet source.

There are numerous other systems in which multi-component outflows are identified. Some can be modelled as a disc wind and a stellar jet; there is no substantial evidence for both these sources to be centred on  distinct stars or for there to be a passive binary companion.
For example, the fast HH\,158 jet originating from DG\,Tau is accompanied by a slow molecular outflow  \citep{2018A&A...620L...1G}. These may share a common origin commensurate with the onion-like layered radial velocity structure \citep{2000ApJ...537L..49B}.  

On the other hand, T\,Tau S is established as a binary, Sa/Sb, with a suggested circumbinary disc and up to three distinct outflows
\citep{ 2005ApJ...628..832D,2016A&A...593A..50K,2018ApJ...861..133Y}. This prompts the question of how these outflows may interact on the scale of their development between 10\,AU and 100\,AU.

\begin{figure}
\centering
\includegraphics[width=0.96\linewidth]{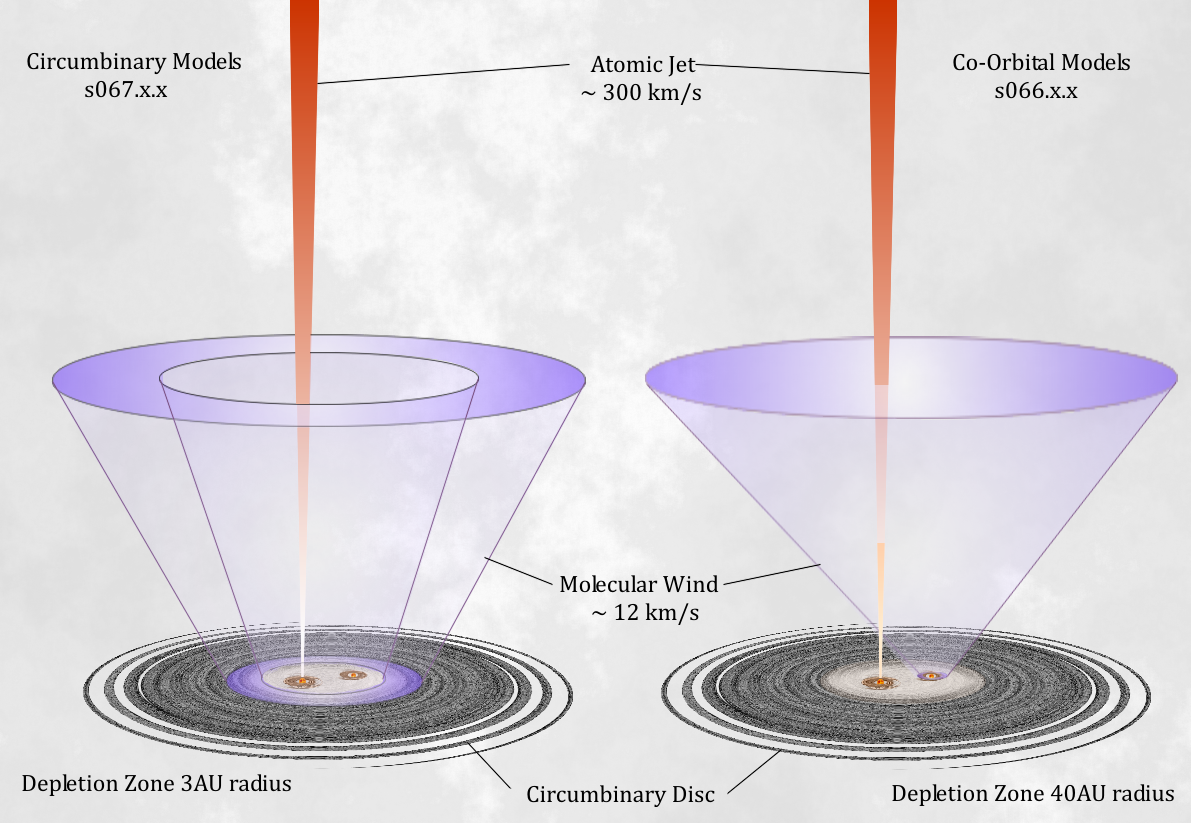}
\caption[Schematic of Circumbinary vs Co-Orbital Models]{A schematic diagram of circumbinary (s067.x.x) vs co-orbital (s066.x.x) models, not to scale.  A `control' simulation within each model series is performed with the atomic jet only -- these are simulations s067.3.3 and s066.2.4 using the terminology introduced in Section~\ref{methods}.}
\label{modelschema}
\end{figure}

Jet simulations provide an interface between physical theory and observation \citep{2014A&A...562A.117T}. 
 Thus we consider here how to simulate  jets from orbiting young stellar objects. 
Two competing scenarios, shown in Fig.\,\ref{modelschema}, are investigated in which the launch site of the molecular
outflow differs. In the Co-orbital Scenario, the molecular flow is launched from the
secondary binary partner, as proposed by \citet{2008MNRAS.387.1313T}.
On the other hand,  in the Circumbinary Scenario advanced by \citet{2012AJ....144...61E}, the jet launches is from the inner edge
of the circumbinary disc. ALMA data for a few sources have been interpreted as favouring this second scenario provided a steady disc wind is assumed \citep{2018A&A...618A.120L,2019ApJ...871..221M}.

\begin{figure}[!ht]
\centering
\begin{center}
\includegraphics[width=0.45\textwidth]{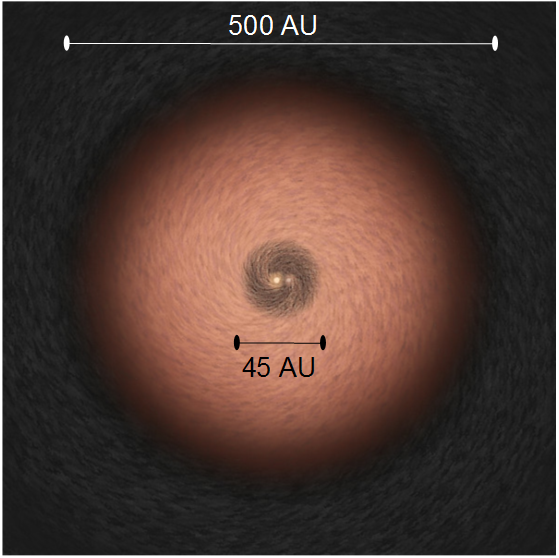}
\end{center}
\caption[HH30 Binary System Schematic]{The underlying YSO system taken here, based on based on HH\,30 with a circumbinary accretion disc, viewed perpendicular to the disc plane; schematic diagram only, central objects not to scale.}
\label{hh30-circumbinary}
\end{figure}

The binary orbit and inner depletion zone of the circumbinary
disc differ between the scenarios as illustrated in Fig.\,\ref{hh30-circumbinary}. In both cases a velocity-pulsed atomic jet emerges from
the more massive object in the binary system. Control simulations were also carried out
in which only the atomic jet was present. 
 
The above scenarios are inspired by the   HH\,30 system which exhibits a well-collimated plume of
hot, optically-emitting atomic and partially ionised hydrogen \citep{2018RMxAA..54..317G} 
and also a colder, dense, wide-angle molecular hydrogen outflow \citep{2007ApJ...660..426H}. The HH\,30 HST jet 
alone has attracted much attention which involves variants on the above models.
 A two-component atomic MHD model was investigated by \citet{2014A&A...562A.117T} employing 
 cylindrical symmetry with a single star-disc outflow system. 
 \citet{2015MNRAS.446.3975S} also investigated a two component system  which results after launch from a Keplerian disc. 
 
 Three dimensional simulations are required due to the two jet sources.  
 Such simulations were performed by \citet{2008A&A...478..453M} by taking two side-by-side nozzles through  
 which atomic jets were injected.  
 They found  that the two jets interfere so that one ends up  engulfed into the second one. In this manner, one can account for
 the propensity of single jet detections despite the high probability for stars to form in binaries.
 They were able to model  L1551 features such as the bending of the secondary jet.
 The innovative work of \citet{2008A&A...478..453M} included attempts to implement a toroidal ambient magnetic field and also to 
 show that slowly orbiting jets, rather than fixed  nozzles, would not alter their high-speed flows which remain controlled by their dynamics. 

The origins for the putative three outflows are assumed to be distributed between the three supplying discs since  circumstellar discs are    invariably associated with collimated outflows. The established mechanism for producing bipolar outflows involves a centrifugally-driven wind flowing along a magnetic field which threads the disc \citep{1982MNRAS.199..883B}. The disc can supply the material for the wind while the wind extracts most of the angular momentum from the disc.  One alternative is the magnetic tower model in which a tightly coiled magnetic field rises from the spinning disc 
 \citep{1985PASJ...37...31S,1996MNRAS.279..389L}. This tower is built on the high magnetic pressure and the steep pressure gradient along the symmetry axis \citep{2012ApJ...757...66H}. A particular mechanism to drive fast atomic jets is based on the magnetic connection between the star and the inner edge of the disc. This is the X-wind model \citep{1988ApJ...328L..19S, 1994ApJ...429..781S}. In the context of this theory, we only consider here a uniform non-rotating supersonic flow from a moving nozzle. We assume that the nozzle scale is sufficient for centrifugal forces to be negligible.

If most stars form in binaries from accretion discs, we should be observing the effects of their combined dual outflows. However,
 simulations of orbiting atomic and molecular outflows have still to be performed. The fast atomic jet may be well collimated while the 
 slow molecular flow may have a large opening angle. In this series of papers, we investigate the potential scenarios and the expected outcomes.
We wish to determine if the nolecules are entrained or destroyed, if the atomic jet is disrupted and ionised in the impact, how the pulses survive and how the altered structure changes with time during an orbit.
 
 This first study is limited to the comparison of a dual co-orbital jet-outflow scenario with a 
  single orbiting  jet  within $\sim$ 120\,AU of the outflow source(s). 
 The simulations were generated using the well-established Eulerian astrophysical code ZEUS-MP. 
 In a second paper, we study and compare the precession-driven (circumbinary) scenario. 
 
 The   physical and numerical assumptions made here are described in Section~\ref{methods} below. We then describe the numerical experiments and the resulting physical structures, especially identifying those which differentiate the scenarios. Specific H$\alpha$ images and CO maps
 highlight the most distinctive signatures. In a further  paper, we will present the complete synthetic observations in terms of a range of imaging and spectroscopic diagnostics.
 
\section{Methods}  
\label{methods}

\subsection{Physics}   

The  equations of fluid dynamics are solved numerically by the standard implementation of the ZEUS-MP astrophysics code  \citep{2006ApJS..165..188H}.  Compressible, non-viscous, non-thermally-conducting flow is assumed; in addition, there are equations for the evolution of radiation energy density and magnetic flux density (the Ideal-MHD approximation).  These couple to the transport equations for momentum and energy by the inclusion of suitable source terms on the right hand sides of these equations.

Energetic collisions lead to dissociation and ionisation of molecules and atoms in addition to electronic, rotational and vibrational transitions. Cooling occurs through radiative transitions, dissociation and via dust grains. We use a modified implementation of ZEUS-MP that contains a simple three-species hydrogen chemistry model that tracks H$_2$, H and H$^+$, based on the model presented in \citet{2007ApJS..169..239G}. Options are implemented in the code for other `zoos' of chemical species, but the three-species chemistry was considered sufficient for our purposes, as the emission lines for our synthetic images (to follow) could all be computed in post-processing either directly from hydrogen species populations or by proxies based on those populations, and more complex chemistry would increase run times and storage requirements.   

To model the cooling of the gas, we use cooling functions that were initially developed by \citet{1997A&A...318..595S} and \citet{2002MNRAS.337..477P} and radically improved by \citet{2007ApJS..169..239G}.  Cooling and chemistry are solved simultaneously for consistency, using an implicit approach to ensure numerical stability. Compressional PdV heating is accounted for separately from radiative heating and cooling using the standard ZEUS-MP approach. The temperature exponents for the cooling terms are pre-tabulated to improve performance, and two separate temperature regimes are identified which are handled by different routines, $T < 300$\,K and $T > 300$\,K, since cooling efficiency is dominated by different components in each regime.  

Though only material phases of H are dynamically traced in the simulations, the model assumes a composition including He and other elements at a concentration typical of the interstellar medium for chemistry and cooling purposes.  This assumed composition permits calculation by proxy of these concentrations when determining synthetic emission properties. 

The combination of orbital motion and jet magnetic field is not beyond the scope of the present study but the results were very limited.
  Firstly, the combination of a  poloidal field and orbital motion was not handled well by the code.  A toroidal field was made to work with orbiting jets, but the main effect was to modify the advancing bow shock into a conical morphology before exiting the grid, with little effect on the behaviour of the jet column.  Both non-zero divergence and tiling issues were encountered. We therefore continued the study omitting the field and so all of the results presented here are derived from hydrodynamical simulations.

Nevertheless, magnetic fields must play an important role in the launch and collimation of protostellar jets with  several basic 
mechanisms available \citep{1982MNRAS.199..883B,1985PASJ...37...31S,1994ApJ...429..781S,2012ApJ...757...66H}. 
Purely radiation-driven or hydrodynamic models are problematic. In radiation models, photon momenta are 
insufficient to drive outflows even from the most luminous sources; while hydrodynamic models require the presence
of a ``nozzle'' formed by pressure gradients in the ambient medium (a De Laval nozzle)
in order to collimate and accelerate a jet. But ambient pressures are not sufficient to
provide this, nor could such a structure resist break up from fluid dynamic instabilities.
Conversely, magnetohydrodynamic  launch models are able to account for both jet launching and collimation. 

Finally, gravitational effects which may change the structure of the jet and ambient media are also omitted. While crucial in the initial acceleration region, we assume that the influence of gravity is negligible  in the propagation region on the short dynamical timescales considered here.

\subsection{Dynamics of the HH\,30 jets}  

 The HH\,30 outflow is thought to be powered  two objects. Estimates put the total mass at $\sim$ 0.45 M$_\odot$ , with typical primary and secondary masses of $\sim$ 0.31$_\odot$ and $\sim$ 0.14 M$_\odot$, though these numbers vary depending on which model is chosen to explain the wiggling behaviour of the knots of bright emission in the jet and the parameter space of each model allows a range of 
 solutions  \citep{2007AJ....133.2799A}.  
  
  Based on these numbers, however, the mean orbital separation is 18\,AU, and the orbital period, 114 years.  The co-orbiting objects are surrounded by a circumbinary accretion disc whose optically illuminated region spans a 500\,AU diameter, with observations in molecular lines suggesting an extended diameter of $\sim$ 850\,AU  \citep{2006A&A...458..841P}.  
\citet{2012AJ....144...61E} find that the binary components orbit within a depletion zone of $\sim$ 40\,AU diameter, and that the orbital motion is the primary driving influence causing the helical wiggling appearance of the jet on parsec  scales, with precession acting as a secondary influence, if present at all. 

To summarise, deduced parameters for the different interpretations are accrued in Table \ref{tab:hh30params}. There are many missing entries where a model did not specify a value.

\begin{table*}
  \caption{Observed parameters of the HH30 system as deduced from several interpretations. No entry means that the value was not specified.}
    \begin{tabular}{lrrrrrl}
    \hline
          & Pety&Anglada& Hartigan& De Colle& Estalella&  \\
          & 2006 & 2007 & 2007 & 2010 & 2012 &  \\
    \hline
    Velocity (Systemic) & 7.25$\pm$0.04 &       &       &       &       & km~s$^{-1}$ \\
    CB Disc Outer Radius & 420$\pm$25 &       &       &       & $\sim$ 250 & AU \\
    CB Disc Inner Radius &       &       &       &       & $\sim$ 40 &  \\
    CS Disc Outer Radius &       &       &       &       & $\lesssim$ 6.00 &  \\
    CS Disc Inner Radius &       &       &       &       & $\sim$ 0.07 &  \\
    Disc Axis Position Angle & 32$\pm$2 & 31.6  &       &       &       & deg \\
    Disc Inclination Angle & 84$\pm$3 &       &       &       &       & deg \\
    Disc Rotation Vector & North-East &       &       &       &       &  \\
    Disc Temperature & 12    &       &       &       &       & K \\
    Precession Angle (A) [2] &       & 1.42$\pm$0.12 &       &       &       & deg \\
    Precession Period (A) [2] &       & 53$\pm$15 &       &       &       & yrs \\
    Half Opening Angle (A) &       & 1.43$\pm$0.12 & 2.6$\pm$0.4 & 2.4   &       & deg \\
    Half Opening Angle (M) & 30$\pm$2 &       &       &       &       & deg \\
    Binary Separation [1] &       & 9-18  &       &       & 18$\pm$0.6 & AU \\
    Binary Separation [2] &       & $< 1$   &       &       &       &  \\
    Absolute Orbit (P) [1] &       &       &       &       & 5.7$\pm$0.9 & AU \\
    Orbit Period [1] &       & 53 &       &       & 114$\pm$2 & yrs \\
    Orbit Period [2] &       & $< 1$ &       &       &       & yrs \\
    Orbital Phase Angle (P) &       &       &       &       & 95$\pm$11 & deg \\
    Orbital Velocity (P) [1] &       &       &       &       & 1.5$\pm$0.2 & km~s$^{-1}$ \\
    Orbital Velocity (S) [1] &       & 2-5   &       &       & $\pm$ & km~s$^{-1}$ \\
    Flow Source (A) [1] &       & Secondary &       &       & Primary &  \\
    Flow Source (A) [2] &       & Primary &       &       &       &  \\
    Flow Velocity, Radial (A) & 200$\pm$0.09 & 100-300 &       &       &       & km~s$^{-1}$ \\
    Flow Velocity, Radial (M) & 12$\pm$2 &       &       &       &       & km~s$^{-1}$ \\
    Flow Velocity, Azim. (A) &       &       &       &       &       & km~s$^{-1}$ \\
    Flow Velocity, Azim. (M) & $< 1.00$ &       &       &       &       & km~s$^{-1}$ \\
    Flow Velocity Variability &       &       &       &       &       &  \\
    Flow Inclination, North (A) &       &       &       & $\sim$ 1 & 5     & deg \\
    Flow Mass (A) & 2 $\times$ 10$^{-8}$ &       &       &       &       & M$_{\odot}$ \\
    Flow Mass (M) & 2 $\times$ 10$^{-5}$ &       &       &       &       & M$_{\odot}$ \\
    Flow Mass Flux (A) & 1 $\times$ 10$^{-9}$ &       &       & 1 $\times$ 10$^{-8}$ &       & M$_{\odot}$ yr$^{-1}$ \\
    Flow Mass Flux (M) & 6.3 $\times$ 10$^{-8}$ &       &       &       &       & M$_{\odot}$ yr$^{-1}$ \\
    Flow Momentum (A) & 4 $\times$ 10$^{-6}$ &       &       &       &       & M$_{\odot}$ km s$^{-1t}$ \\
    Flow Momentum (M) & 2.4 $\times$ 10$^{-4}$ &       &       &       &       & M$_{\odot}$ km s$^{-1t}$ \\
    Flow Momentum Flux (A) & 2.6 $\times$ 10$^{-7}$ &       &       &       &       & M$_{\odot}$ km s$^{-1}$ / yr \\
    Flow Momentum Flux (M) & 7.5 $\times$ 10$^{-7}$ &       &       &       &       & M$_{\odot}$ km s$^{-1}$ / yr \\
    Flow Ionisation (A) &       &       & 0.05-0.40 &       &       &  \\
    Flow Temperature (A) &       &       & 7.26$\times$10$^3$&$\sim$ 1$\times$10$^4$&       & K \\
    Flow No. Density (A) &       &       & 1$\times$10$^6$ &       &       &  \\
    Flow Width (A) @ 20 AU &       &       & 14$\pm$3 & 15    &       & AU \\
    Flow Width (A) @ 500 AU &       &       & 36$\pm$4 & 30    &       & AU \\
    Stellar Mass (Total) [1] & 0.45$\pm$0.04 & 0.25-2 &       &       & 0.45$\pm$0.04 & M$_{\odot}$ \\
    Stellar Mass (P) [1] &       & 0.25-1 &       &       & 0.31$\pm$0.04 & M$_{\odot}$ \\
    Stellar Mass (S) [1] &       &       &       &       & 0.14$\pm$0.03 & M$_{\odot}$ \\
    Stellar Mass (P) [2] &       & 0.1-1 &       &       &       & M$_{\odot}$ \\
    Stellar Mass (S) [2] &       & 0.01-0.04 &       &       &       & M$_{\odot}$ \\
    Stellar Luminosity (Total) & 0.2   &       &       &       &       & L$_{\odot}$ \\
    \hline
\multicolumn{7}{p{0.76\linewidth}}{Summarises the findings of a number of observations and investigations into the nature of HH\,30.  
This is not intended to be exhaustive but provides the basis for our choice of model parameters. 
(A) designates a parameter relating to the atomic jet; (M) to the molecular outflow. 
 [1] and [2] are alternative scenarios - [1] is orbital, [2] is precession.  
 Any parameter not identified as either is agnostic or else, by default, assumes the orbital scenario.}
    \end{tabular}%
  \label{tab:hh30params}%
\end{table*}%

It has been suggested that there is variability in the jet velocity on long and short periods. The longer-period variability in ejection velocity is likely to give rise to the knots of bright emission found in the HH\,30 outflow on the 0.1\,parsec  scale \citep{1990ApJ...364..601R}.  The short-period variability in the ejection velocity ($\sim$ months), possibly chaotic in nature, arises from variable accretion, which steepens into shock fronts that provide the main cause of heating and ionisation of the jet material \citep{2007ApJ...660..426H,2007AJ....133.2799A}.
In addition, the parsec scale outflow of HH\,30 appears to be driven sideways in the sky plane, exhibiting a West-facing 'C' shape curvature that may arise from systemic velocity, or impinging outflows or winds from nearby objects \citep{2012AJ....144...61E}.

A parallel strand of investigation into the HH\,30 system involved observation in molecular lines \citep[e.g.][]{2006A&A...458..841P}. Mapping of HH\,30 in HCO$^+$ and several isotopologues of the CO molecule which emit at milimetre wavelengths.  
reveals a great deal, including Keplerian rotation of the accretion disc in the $^{13}$CO(2-1) 1.35mm line. Of particular interest for our work is the observation of a slower-moving, cold, dense outflow imaged in the $^{13}$CO(2-1) 1.35mm line.  This outflow of $\sim$ 12 km\.s$^{-1}$ takes the form of a wide plume and  is only observed emerging from the north-facing side of the accretion disc, where the atomic outflow is also most active, and is quiescent on the south face, similar to the atomic counterjet.

Later work by \citet{2008MNRAS.387.1313T} goes a stage further in developing a working model of the HH\,30 molecular outflow, by performing simulations in which the outflow material is composed of particles that are ejected ballistically into the problem domain.  Four scenarios are investigated, based on the work of \citet{2007AJ....133.2799A}. The first three scenarios assume orbital motion of the molecular outflow source, with varying parameterization.  The fourth scenario assumes that the binary system is very close and that the wiggle in the atomic jet arises from tidally-induced precession.  In this scenario, the molecular outflow originates from the circumbinary disc, whose inner radius is much smaller (3\,AU) because of the tight binary orbit (0.75\,AU).

\citet{2008MNRAS.387.1313T}  find that the model which assumes a very close binary orbit and a circumbinary disc source produces a closer resemblance to the observed morphology of the molecular outflow.  This implies that precession of the atomic jet source is the primary origin of jet wiggling, with orbital motion a lesser influence. 

In summary, analytical modelling of the atomic jet wiggle has suggested that the first scenario is the more likely, with the molecular outflow therefore being produced by an orbiting binary partner also.  However, computational modelling of the behaviour of the molecular outflow in the two cases favours the precession-driven scenario.

\begin{figure*}
\centering
\captionsetup[subfloat]{justification=centering}
\subfloat[Simulation Time: 63 years]{\includegraphics[width=0.96\linewidth]{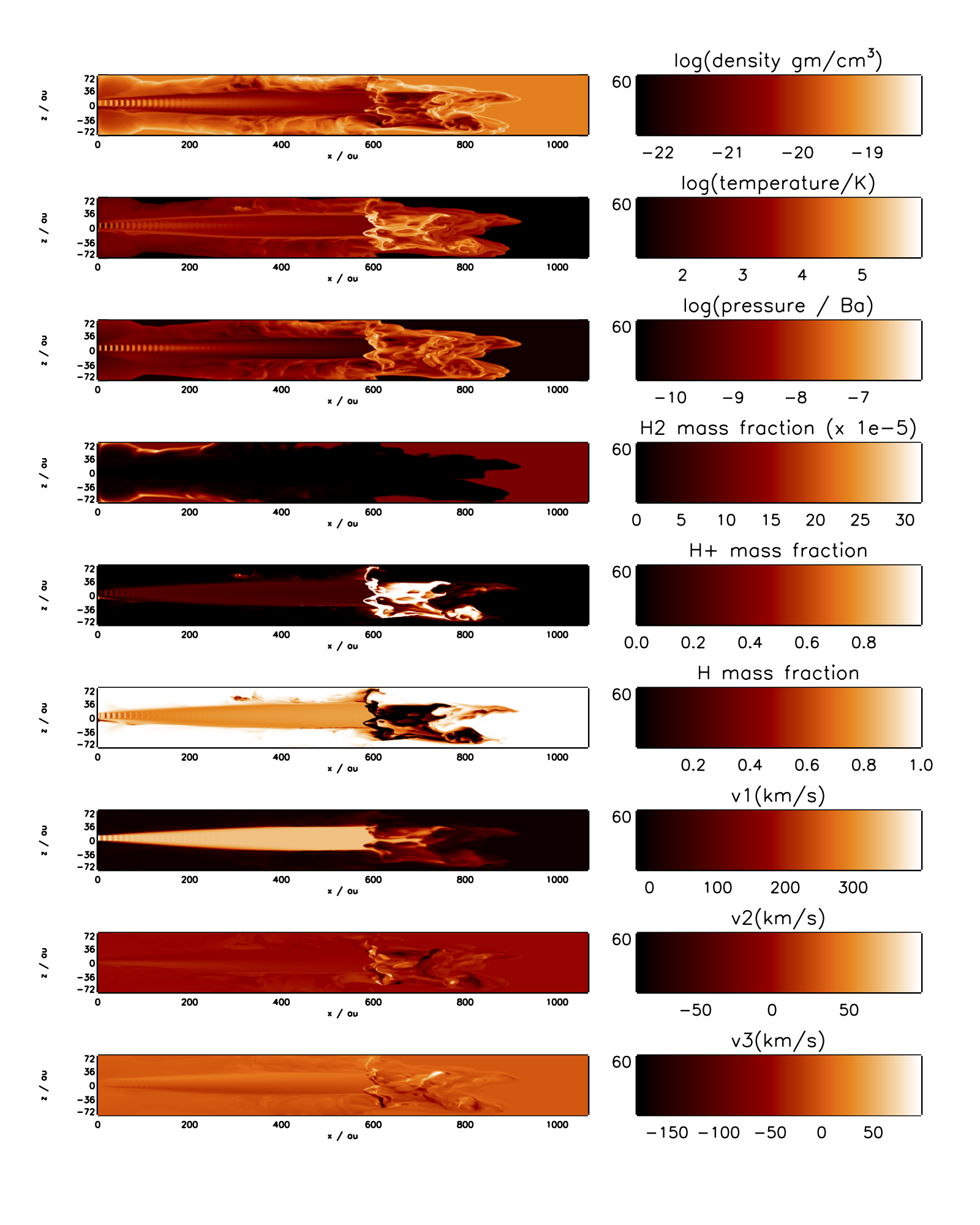}}
\caption[Prototyping the HH\,30 Atomic Jet: Physical Variables, model s064.2.2]{Prototyping the HH\,30 atomic jet: physical variables at the early time of 63 years, model designated s064.2.2.  This jet is from an orbiting inlet, the motion of which resembles the co-orbital series of models (s066.x.x) in the followingdiscussion. The panels show the $x$-$z$ mid-plane with the scale in AU.}
\label{proto1}
\end{figure*} 

\begin{figure*}
\centering
\captionsetup[subfloat]{justification=centering}
\subfloat[Simulation Time: 189 years]{\includegraphics[width=0.96\linewidth]{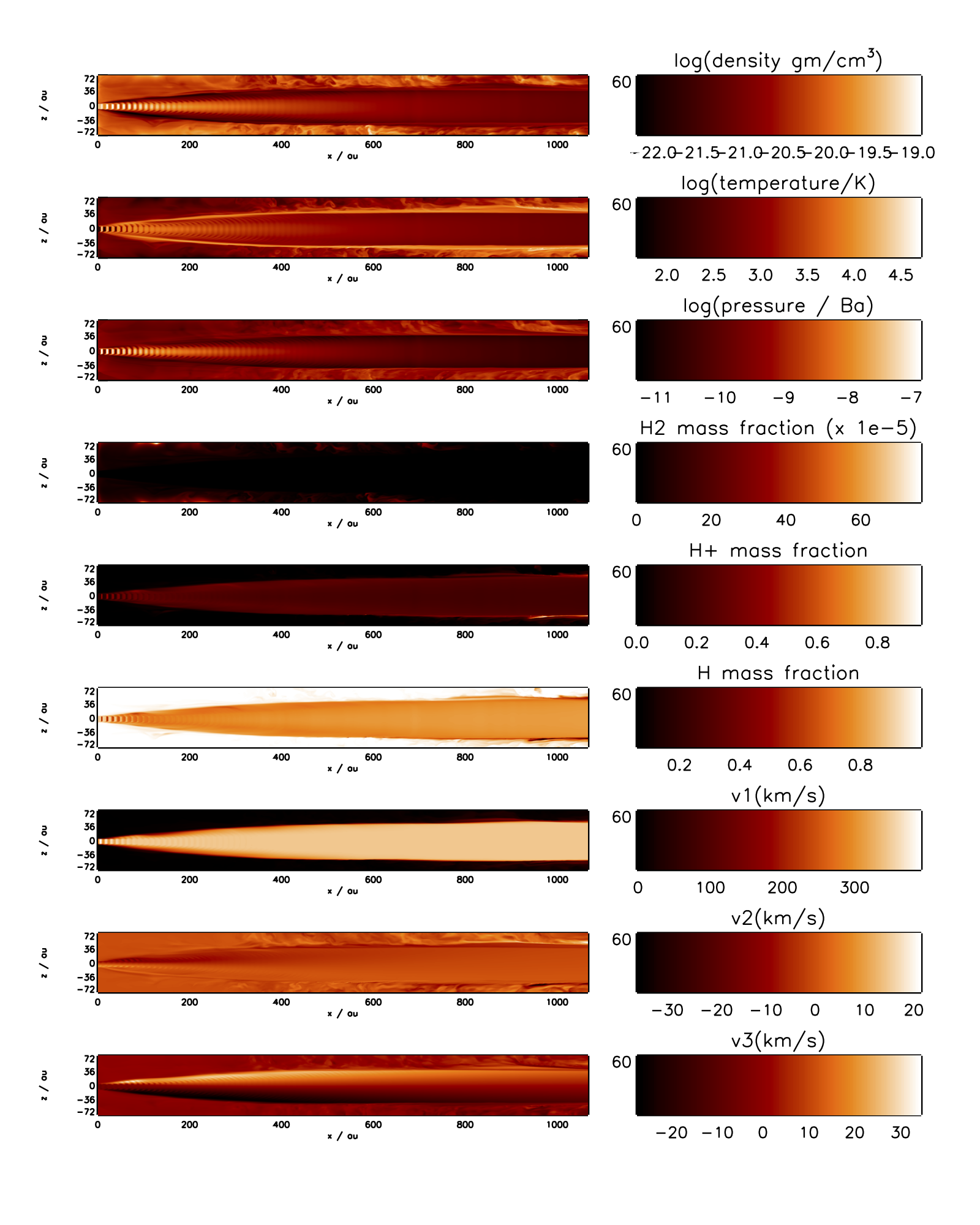}}
\caption[Prototyping the HH\,30 Atomic Jet: Physical Variables, model s064.2.2]{Prototyping the HH\,30 atomic jet: physical variabls at the 
late time of 189 years, model  designated s064.2.2.  This jet is from an orbiting inlet, the motion of which resembles the co-orbital series of models (s066.x.x) discussed below. The panels show the $x$-$z$ mid-plane with the scale in AU.}
\label{proto2}
\end{figure*}

\subsection{Implementing an Orbiting Source}
In this paper, we present results from models of protostellar jets that are launched from a source undergoing orbital motion as part of a binary system.  Numerical models of protostellar jets often assume a frame of reference in which the jet/counterjet source is at rest with respect to the surrounding medium, or else in a state of rectilinear motion.  However, from $N$-body simulations of star cluster formation, and from our physical observations of the end products of star formation, these sources are often expected to occur in binary systems and will therefore be subject to acceleration.  In addition, axisymmetric approximations are often imposed to reduce computational demands and to focus attention on a particular piece of internal jet physics, but in the case of jets from an orbiting source, this is not possible and a full three-dimensional model must be computed. 

We assume here  a binary protostellar system that possesses sufficient orbital stability to be well-described by Kepler's equations.  In reality, the objects   are still in the process of accretion from a surrounding circumbinary disc of infalling material and so their mass will be time-varying and hence their orbital characteristics will not be constant.  However, the time scales of the simulated jets are small in comparison to the accretion timescale and so the assumption of constant mass is a reasonable one.  
 
 It is also possible that the orbits might be perturbed by anisotropy in the gravitational attraction of the surrounding accretion disc if there are local variations in the amount of matter in the disc.  We assume that any accretion disc clumpiness is not biased significantly in any direction and thus any such perturbations may be ignored. 
 
Kepler;s equations describe the elliptical orbit of a member of a stable binary star system.  In such a system the binary partners co-orbit a common centre. This barycentre lies at one of the focii of each elliptical orbit.  Depending on the mass ratio of the partners the barycentre may be well outside both objects, but might also be embedded within the more massive one.

The code takes the orbital period, $T$, as given by  $T^2 = {4 \pi^{2} / (a^{3}}{G(m_1 + m_2)})$ where $a$ is the 
elliptical semi-major axis between masses $m_1$ and $m_2$. 
 The relationship between eccentric anomaly, $E$, and mean anomaly, $M(t)$, is
\begin{equation}
\label{eanom}
    M(t) = \frac{2 \pi t}{T} = E - \epsilon \cdot \sin(E),
\end{equation}
where $\epsilon$ is the orbital eccentricity parameter.

The relationship between true anomaly, $\theta$, and eccentric anomaly is
\begin{equation}
\label{ctanom}
   \cos (\theta) = \frac{\cos (E) - \epsilon}{1 - \epsilon \cdot \cos (E)} ,
\end{equation}
with the  radial distance from the barycentre as a function of the true anomaly given by
$r(\theta)$ = 
${a(1-\epsilon^{2})}/(1 + \epsilon \cos(\theta))$.
In our implementation, the centre of the jet inlet on the $x=0$ boundary is repositioned with each timestep, with the motion corresponding to that described by the Kepler equation.  The key to this is determining the eccentric anomaly, $E$.  Rearranging into a non time-dependent form, we obtain:
\begin{equation}
f(E) = 0 = E - \epsilon \cdot sin(E) - \frac{2 \pi t_0}{T},
\end{equation}
where $t_0$ is a specified, and thus constant, interval of time.  This is a transcendental equation, the roots of which cannot be determined analytically. Therefore, a numerical method is required.  In our implementation the Newton-Raphson method is used, which converges to 8 decimal places of accuracy after 5 iterations:
\begin{equation}
E_{n+1} = E_n - \frac{E_n - \epsilon \cdot \sin(E_n) - E_0}{1 - \epsilon \cdot \cos(E_n)}.
\label{newtonkepler}
\end{equation}
in which the initial guess $E_0$ is taken to be the mean anomaly $\frac{2 \pi t_0}{T}$. Given an accurate estimate of the eccentric anomaly, we can then calculate the true anomaly and the radial distance from the barycentre then follows. It is then simple trigonometry to calculate where the centre of the jet inlet must lie.

\section{Results}  
 
 \subsection{Orbiting atomic jet on 1,000\,AU scale}  
 
We first simulate an extended atomic jet from an orbiting young star with the same parameters as the co-orbiting jet runs  (see Table \ref{tab:hh30params}). 
This run with a single orbiting jet permits us to calibrate the code and check the dynamical behaviour.
 Figures \ref{proto1} \& \ref{proto2} display all the physical outputs from this simulation at an early time of 63 years and a late time of 189 years;
 respectively.     In the absence of a wide-angle molecular flow, a long problem domain is feasible; a 1,280 $\times$ 175 $\times$ 175 mesh was employed, running on 160 cores.  The jet was 10$\times$ over-pressured with respect to the ambient medium and its inlet radius was 7\,AU. 
 
 The panels of  Figures \ref{proto1} \& \ref{proto2} show all the expected properties at an early time,with the leading bow on the domain,
 and at a late time when the flow has settled. A low-density low-pressure cavity is formed around the jet. Energy is transferred into the ambient medium which displays strong turbulence. The jet pulses steepen abruptly into shocks but the shocked layers expand as the shocks weaken, with the layers merging at $\sim$ 200\,AU.
 A small fraction of  molecular hydrogen is seen to form in the compressed ambient medium. The initial region of high ionisation is advected downstream with the advancing bow shock leaving only low ionisation regions within the jet and cavity.

\subsection{Dual outflows}  
 
 Although guided by HH\,30, we are interested in the general problem as illustrated in  Fig.\,\ref{modelschema}. To quantify, a wide range of simulations were performed until a small number of relevant conditions could be  taken forward for close study.
 
 The molecular outflow was added to the fast atomic jet through a distinct module. There was no method that could be found to sensibly implement a boundary condition where the two outflows were already interacting prior to their incursion into the domain.
  
  Table \ref{tab:modelgeometry} summarises the problem domain setup for ZEUS-MP used in the models that follow. A stack in the x-dimension of 20 thin slabs 8 zones in width, with each slab consisting of a 3 x 3 arrangement of square 115 x 115 zone tiles in the y-z plane was taken.
  This has the advantage of fully containing the flow inlets and their orbits within the central tile.  The 180 core configuration that was settled on, arising from a choice of 20 longitudinal slabs, worked well and queuing times were generally satisfactory.  
  
  Due to the wide angle of the molecular flow, simulations require a model geometry with a large lateral span.  To accommodate an additional 30$^{\circ}$ molecular outflow, without truncating the problem domain, would require greater computing resources than those available.  It was also desirable to increase the spatial resolution to capture the features of the flow in more detail for final results. It may well be that  the most significant results relevant to  future observations will appear within the first 100\,AU of the launch regions.

\begin{table*}
\resizebox{0.9\textwidth}{!}{\begin{minipage}{\textwidth}
  \centering
  \caption{Standard model geometry and tiling}
    \begin{tabular}{crrrrrcr}
    \hline
    Coordinate & Min   & Max   & Span  & Grid  & Zone Size & MPI   & Zones \\
          & (cm)  & (cm)  & (cm)  & Zones &  (cm) & Tiling & /Tile \\
    \hline
    x     & 0     & 1.600E+15 & 1.600E+15 & 160   & 1.0E+13 & 20    & 8 \\
    y     & -1.725E+15 & 1.725E+15 & 3.450E+15 & 345   & 1.0E+13 & 3     & 115 \\
    z     & -1.725E+15 & 1.725E+15 & 3.450E+15 & 345   & 1.0E+13 & 3     & 115 \\
    \hline
    \end{tabular}%
  \label{tab:modelgeometry}%
\end{minipage} }
\end{table*}%

Table \ref{tab:commonparams} summarises the choices of parameters for the outflows and the initial ambient medium which was chosen to be entirely atomic with a temperature of $\sim$ 100\,K.    In fact, the simulations themselves were allowed to run for a sufficient length of time to `nurture' their own problem domains; particularly in the simulation runs where a molecular outflow component was present.
The parameters in the upper section of Table \ref{tab:commonparams} are  configurable  and could be input directly to ZEUS.  The dependent parameters in the lower section are all quantities of interest that are directly calculated.

\begin{table*}
  \caption{Outflow and ambient medium parameters}
  \vspace{0.1em}
    \begin{tabular}{clrrrrrl}
    \hline
          & {Parameter} & {Atomic Outflow} &   & {Molecular Outflow} & & {Atomic} & {Units} \\
       &       & {Type I} & {Type II} & {Type I} & {Type II} & {Medium} &  \\
    \hline
     & Source Object & {Primary} & {Primary} & {Secondary} & {CB Disc} &       &  \\
          & Inner Radius &       &       &       & 1.33E+14 &       & cm \\
          & Outer Radius & 5.50E+13 & 5.50E+13 & 1.20E+14 & 3.20E+14 &       & cm \\
          & Density &       &       &       &       & 1.2525E-18 & g/cm$^3$ \\
          & Energy Density &       &       &       &       & 1.3284E-08 & erg/cm$^3$ \\
          & Density Ratio & 1     & 1     & 50    & 10    &       &  \\
          & Pressure Ratio & 10    & 10    & 7.5   & 1.5   &       &  \\
          & Mach No. & 95    & 95    & 30    & 30    &       &  \\
          & Rotation (Solid) & 1.31E-08 & 1.31E-08 & 6.53E-09 &       &       & rad/s \\
          & Rotation (Kepler) &       &       &       & 3.50E-09 &       & rad/s \\
          & Radius (Kepler) &       &       &       & 2.20E+14 &       & cm \\
          & Orbital Separation & 18    & 0.75  &       &       &       & AU \\
          & Mass (Primary) & 0.31  & 0.44  &       &       &       & M$_{\odot}$ \\
          & Mass (Secondary) & 0.14  & 0.10  &       &       &       & M$_{\odot}$ \\
          & Precession Rate &       & 3.76E-09 &       &       &       & rad/s \\
          & Precession Angle &       & 0.025 &       &       &       & radians \\
    \hline
     & Inner R (zones) &       &       &       & 13.265 &       &  \\
          & Outer R (zones) & 5.5   & 5.5   & 12    & 32    &       &  \\
          & Orbit Period & 3.59E+09 & 3.06E+07 &       &       &       & s \\
          & Precession Period &       & 1.67E+09 &       &       &       & s \\
          & Adiabatic Exponent & 1.66667 & 1.66667 & 1.42857 & 1.42857 & 1.66667 &  \\
          & No. Density & 7.00E+05 & 7.00E+05 & 3.50E+07 & 7.00E+06 & 7.00E+05 &  \\
          & Density & 1.25E-18 & 1.25E-18 & 6.26E-17 & 1.25E-17 & 1.25E-18 & g/cm$^3$ \\
          & Temperature & 1090  & 1090  & 16.4  & 30    & 109   & K \\
          & Sound Speed & 3.43E+05 & 3.43E+05 & 3.89E+04 & 4.20E+04 & 1.09E+05 & cm/s \\
          & Inlet Flow Speed & 3.26E+07 & 3.26E+07 & 1.26E+06 & 1.17E+06 &       & cm/s \\
          & Inlet Area & 9.50E+27 & 9.50E+27 & 4.52E+28 & 2.66E+29 &       & cm$^2$ \\
          & Mass Throughput & 5.10E-09 & 5.10E-09 & 5.64E-08 & 6.15E-08 &       & M$_{\odot}$ / yr \\

    \hline
    \end{tabular}%
  \label{tab:commonparams}%
\end{table*}%


 Table \ref{tab:simulationruns} summarises the characteristics of the simulation runs. Co-orbital runs are designated by the prefix s066 and the circumbinary by s067. Four major long timescale (175 yrs) runs were performed to establish fully developed flow structures. Both models were run with and without the molecular flow, designated with the suffix 2.3 and 2.4, respectively.
 
 Six additional, shorter duration simulations were then carried out for each scenario; three with different velocity pulse time periods for the atomic jet and three with different values of orbital eccentricity.  The data retained for these runs starts from 65 years into the outflow evolution (by which time the outflows have crossed the problem domain and exhibit fully developed flow) and tracks the evolution over a further 22 years.

The velocity pulse period $T_{\rm vpulse}$ determines the period of the time-varying sinusoidal signal imposed on the velocity of injected material.  For all the presented simulations, the Relative Amplitude parameter $A_R$ used  is set at 0.2.  This generates a velocity signal:
\begin{equation}
V_J(t) = M \times C_J \times \frac{1+A_R\ \cos \omega_{v}t}{1+A_R},
\end{equation}

\noindent where $M$ is the Mach number of the jet, $C_J$ is the jet sound speed, and $\omega_{\rm v} = 2\pi/T_{\rm vpulse}$.  The maximum of this signal is $MC_J$ and the minimum is 66\% of this value. The orbital eccentricity parameter, $\epsilon$, is used in Equation \ref{newtonkepler} to compute the Keplerian orbit of the jet inlet by the Newton-Raphson method. The HDF data dumps are produced at simulation time intervals of 
2.125 $\times$ 10$^6$ seconds.  This is chosen to be less than half of $T_{\rm vpulse}$.  It is a standard 
result in signal processing that in order to capture a sinusoidal signal the sampling frequency must be at least twice the highest frequency component of the signal.  With our choice of $T_S \approx 2.5 T_{\rm vpulse}$, `strobe' effects are avoided in animations and the possibility exists to determine proper motions of structures within the flow.

 Storage requirements were considerable.  Each data dump in the listed simulations occupied 799 MB of memory space. Thus, the four 'primary' simulations required over 2 TB of storage each; whilst the twelve shorter simulations required 260.5 GB each.

\begin{table*}
  \centering
  \caption{Simulation Runs}
    \begin{tabular}{rrccrcrrrr}
    \hline
     &   {Simulations} & {Outflows} &     & Vpulse & Orbital &   &   {Retained Dumps} \\
      &   & Atomic & Molecular & Period & $\epsilon$     & First & Last  & T$_1$ (Y) & T$_2$ (Y) \\
    \hline
          &       &       &       &       &       &       &       &       &  \\
     & {s066.2.3} & I     & I     & 5.256E+06 & 0.00  & 1     & 2599  & 0     & 175 \\
          &       &       &       &       &       &       &       &       &  \\
          & \multicolumn{1}{l}{s066.2.4} & I     & -     & 5.256E+06 & 0.00  & 1     & 2599  & 0     & 175 \\
          &       &       &       &       &       &       &       &       &  \\
          & {s066.4.3} & I     & I     & 1.051E+07 & 0.00  & 975   & 1300  & 65    & 87 \\
          & {s066.4.4} & I     & I     & 7.884E+06 & 0.00  & 975   & 1300  & 65    & 87 \\
          & {s066.4.5} & I     & I     & 1.314E+07 & 0.00  & 975   & 1300  & 65    & 87 \\
          &       &       &       &       &       &       &       &       &  \\
          & {s066.5.25} & I     & I     & 5.256E+06 & 0.25  & 975   & 1300  & 65    & 87 \\
          & {s066.5.50} & I     & I     & 5.256E+06 & 0.50  & 975   & 1300  & 65    & 87 \\
          & {s066.5.75} & I     & I     & 5.256E+06 & 0.75  & 975   & 1300  & 65    & 87 \\
          &       &       &       &       &       &       &       &       &  \\
    \hline
          &       &       &       &       &       &       &       &       &  \\
     &{s067.3.2} & II    & II    & 5.256E+06 & 0.00  & 1     & 2599  & 0     & 175 \\
          &       &       &       &       &       &       &       &       &  \\
          & {s067.3.3} & II    & -     & 5.256E+06 & 0.00  & 1     & 2599  & 0     & 175 \\
          &       &       &       &       &       &       &       &       &  \\
          & {s067.4.3} & -     & II    & & -     & 1     & 1199  & 0     & 80 \\
          &       &       &       &       &       &       &       &       &  \\
          & {s067.4.3} & II    & II    & 1.051E+07 & 0.00  & 975   & 1300  & 65    & 87 \\
          & {s067.4.4} & II    & II    & 7.884E+06 & 0.00  & 975   & 1300  & 65    & 87 \\
          & {s067.4.5} & II    & II    & 1.314E+07 & 0.00  & 975   & 1300  & 65    & 87 \\
          &       &       &       &       &       &       &       &       &  \\
          & {s067.5.25} & II    & II    & 5.256E+06 & 0.25  & 975   & 1300  & 65    & 87 \\
          & {s067.5.50} & II    & II    & 5.256E+06 & 0.50  & 975   & 1300  & 65    & 87 \\
          &{s067.5.75} & II    & II    & 5.256E+06 & 0.75  & 975   & 1300  & 65    & 87 \\
          &       &       &       &       &       &       &       &       &  \\
    \hline
    \end{tabular}%
  \label{tab:simulationruns}%
\end{table*}%

\subsection{A single orbiting atomic jet at high resolution}

The s066.x.x series of simulations model a scenario in which the fast-moving, atomic, optically emitting outflow of a T\,Tauri star is launched by the more massive binary partner in a two-star system, and the slow-moving, wide-angle molecular flow is launched from its lower-mass co-orbiting partner.  In this scenario the co-radius is 18\,AU and the masses of the two objects are 0.31\,M$_{\odot}$ and 0.14\,M$_{\odot}$.  Orbital eccentricity is considered in the final three simulations in this set in which values of $\epsilon$ range from 0.25 to 0.75.

\begin{figure}
\subfloat[$z-y$ plane, $x = 0$~cm]{\includegraphics[width=0.49\linewidth]{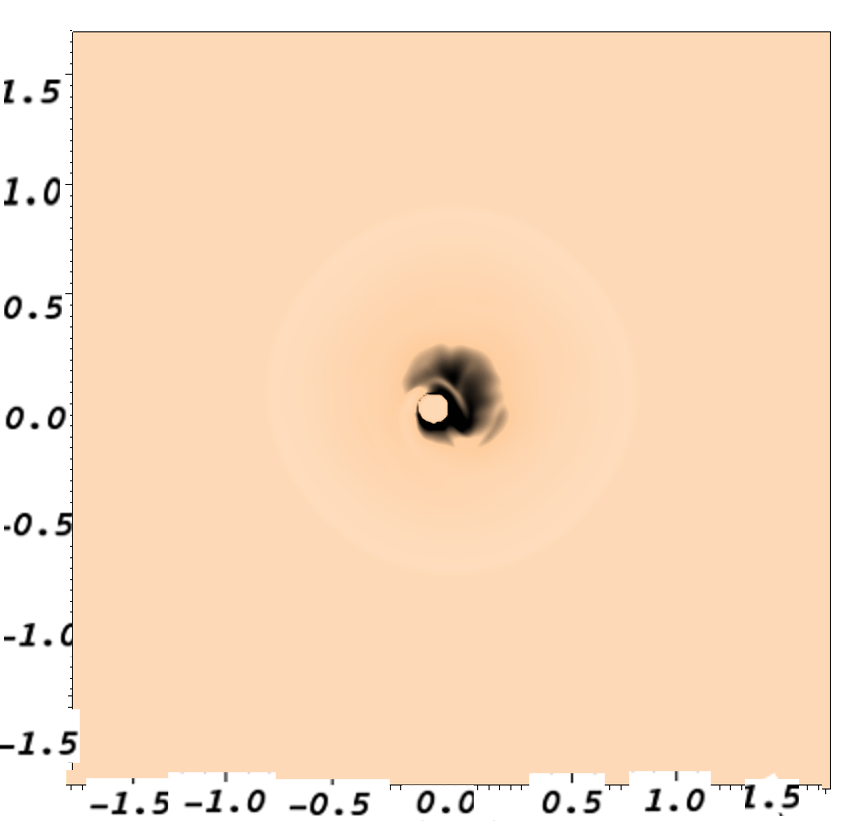}}
\subfloat[$z-y$ plane, $x = 7.5 \times 10^{14}~$cm]{\includegraphics[width=0.49\linewidth]{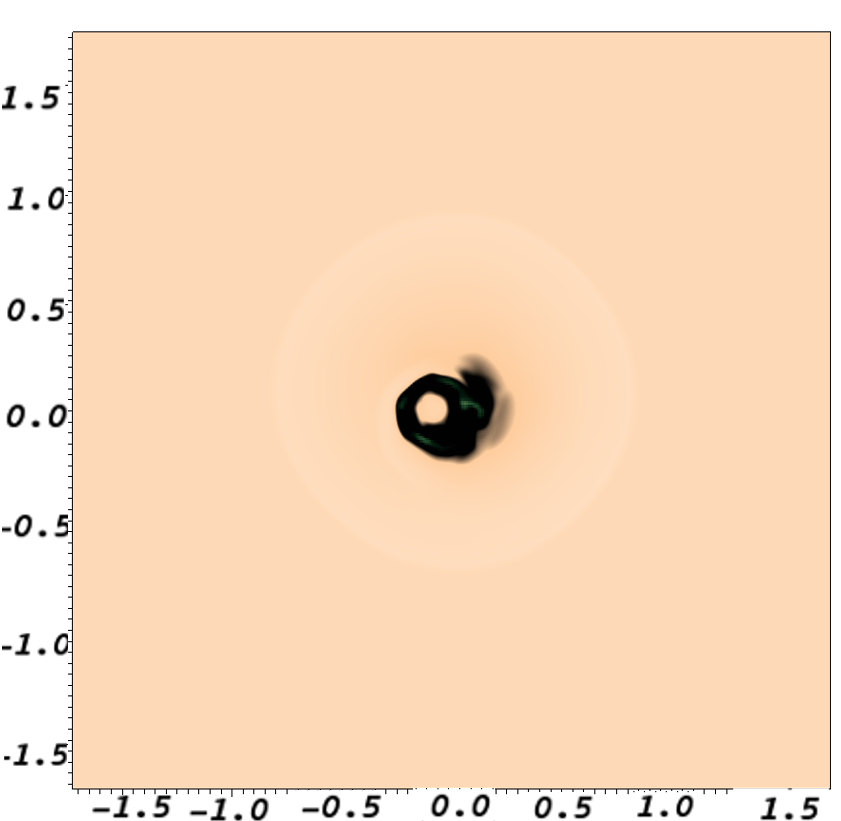}} \\\\
\subfloat[$z-y$ plane, $x = 1.5 \times 10^{15}$cm]{\includegraphics[width=0.49\linewidth]{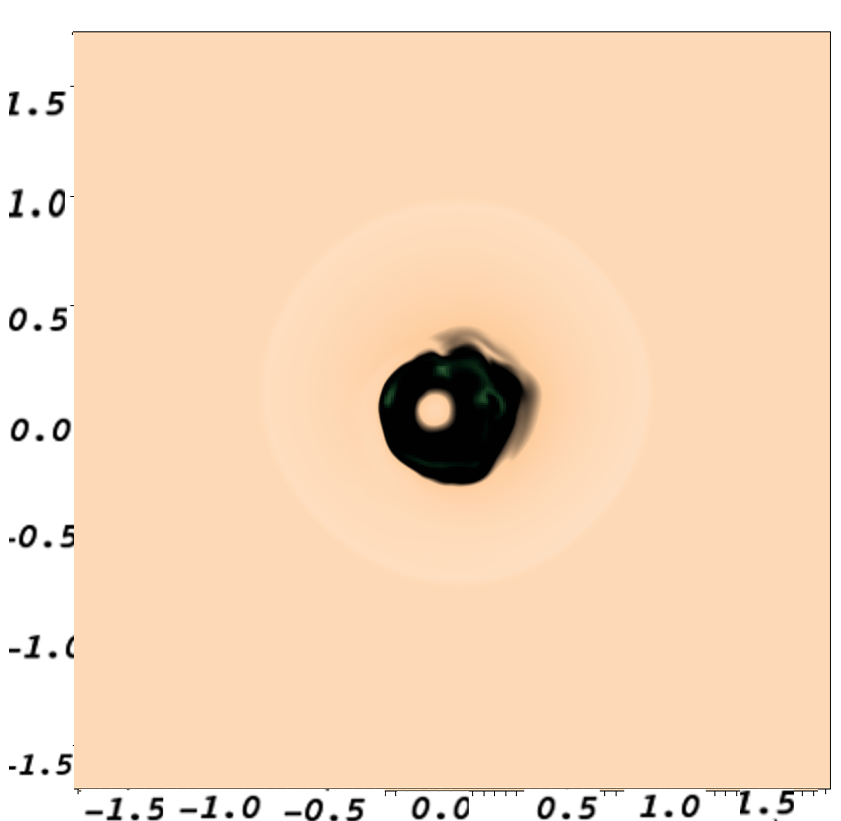}}\hspace*{0.02\linewidth} 
\subfloat[$x-y$ plane, z=0; $x-z$ plane, y = 0]{\includegraphics[width=0.49\linewidth]{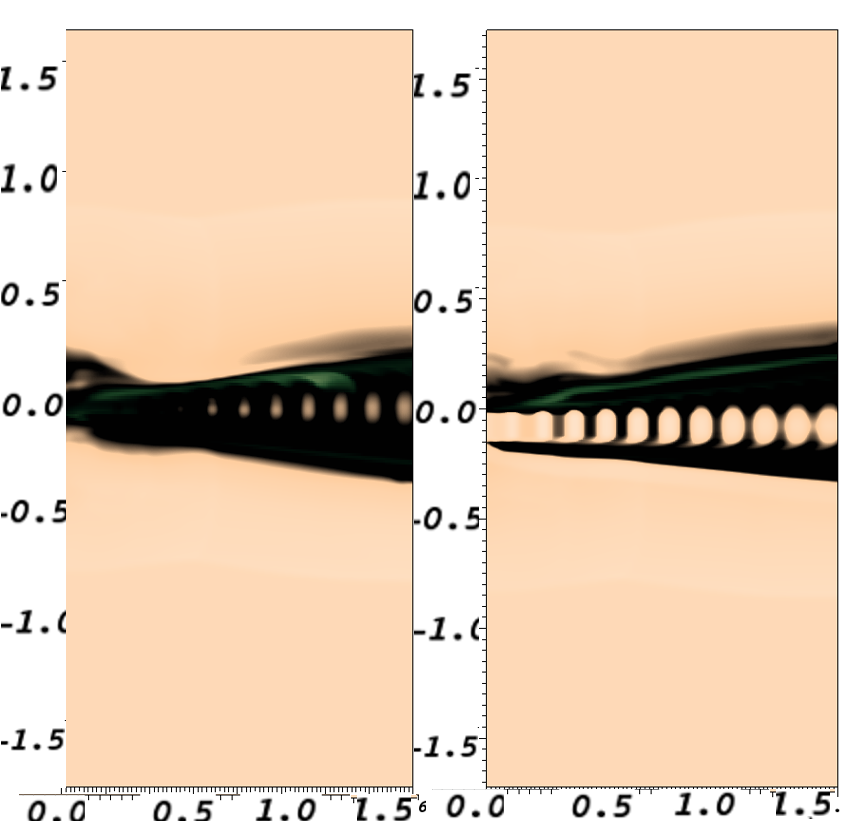}}\\\\
\hspace*{0.6cm} \subfloat[Density (g \, cm$^{-3}$)]{\includegraphics[width=0.30\linewidth]{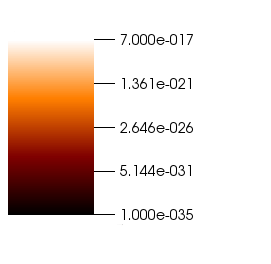}}
\subfloat[H$_2$ density (g \, cm$^{-3}$)]{\includegraphics[width=0.30\linewidth]{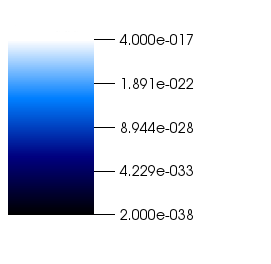}}
\subfloat[Fractional ion.]{\includegraphics[width=0.30\linewidth]{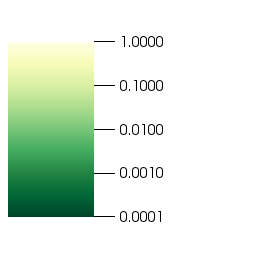}}
\\
\caption[Co-orbital Model s066.2.4: Section Plots, T=87.5 Years]{\textbf{Single atomic model under co-orbital conditions} s066.2.4: 18\,AU binary,  density cross-sections at simulation time 87.5 years.  Axis scales are in units of 10$^{15}$cm.  Ambient medium is atomic  with trace molecular hydrogen formed  during the simulation.  Underlying density plot is fully opaque.  For clarity, H$_2$ and ionisation fraction overplots are at ramped opacity; 100\% opacity at maximum value, transparent at minimum value.}
\label{6624_1300_xsect}
\end{figure}

The analysis of the Single Atomic Outflow Case (simulation s066.2.4) begins from
Figure \ref{6624_1300_xsect} which shows a set of cross-sectional  plots of the case where the molecular outflow is absent.
The simulation time is 87.5 years.  Note that this figure shares a common colour scaling of variables and as a result some features appear faint. 
In spite of this, close examination reveals a spiral pattern in the density which radiates outwards from the jet column, induced in the surrounding medium by the orbital motion.  Panel (a) shows the jet inlet at x=0, and panels (b) and (c) show  cross-sections further along the x-direction.  

In subfigure (d), the jet is seen entering the domain from the left.  The density distribution is sensitive to the jet location relative to the chosen mid-plane at the simulation time. An expanding cocoon of lower density material surrounds the denser jet column.  We also see from panel  (d) that the pulsed velocity signal has given rise to small-scale density knots within the jet column, sandwiched between regions of lower density.  The knots appear to be expanding in the direction of travel as they cross the domain and exit the far x-boundary but remain well-collimated in the y-z plane.
The density knots are a feature of all the models though they become somewhat disrupted in the Co-orbital Mode where the atomic jet is in collision with the molecular outflow.
  
\begin{figure*}
\subfloat[density and composition at x=0\,cm]{\includegraphics[width=0.24\textwidth]{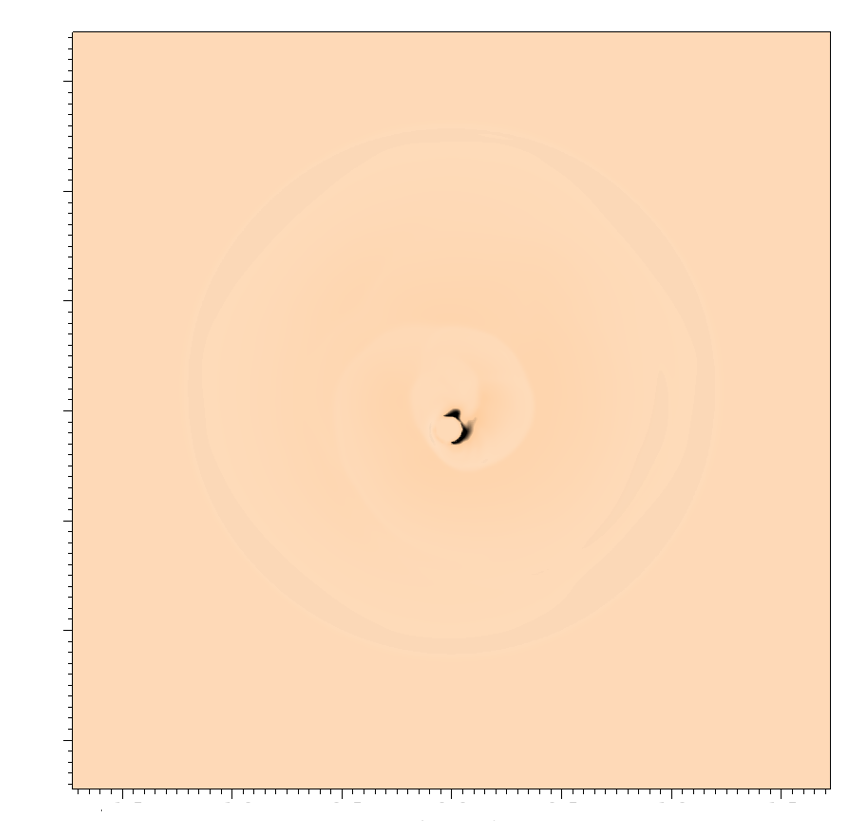}}
\subfloat[density and composition, 50AU]{\includegraphics[width=0.24\textwidth]{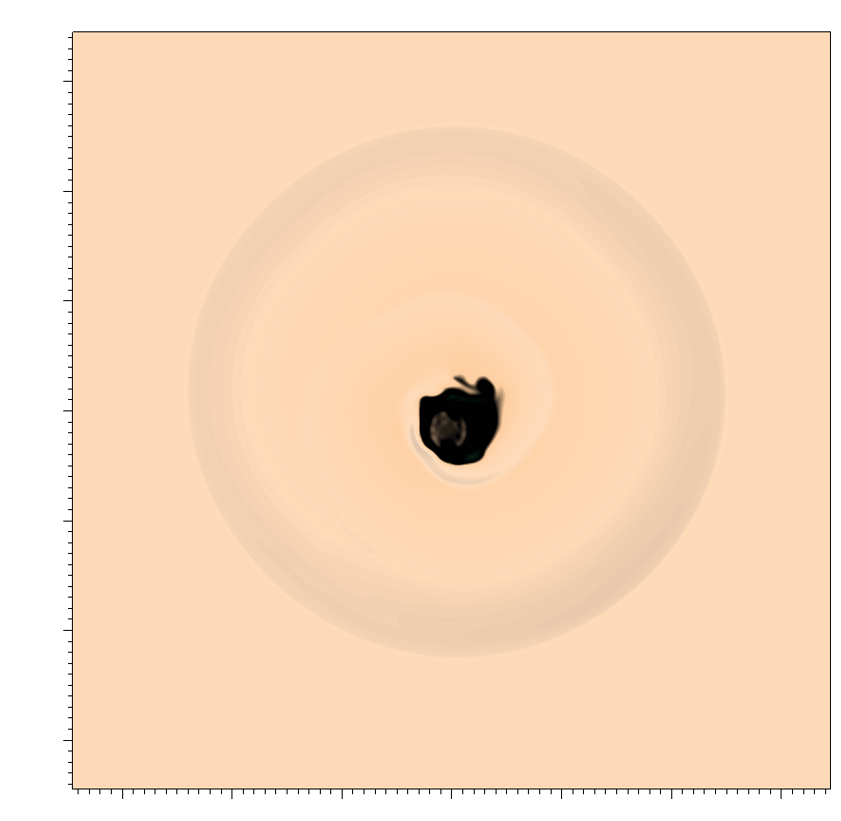}}
\subfloat[density and composition, 100AU]{\includegraphics[width=0.24\textwidth]{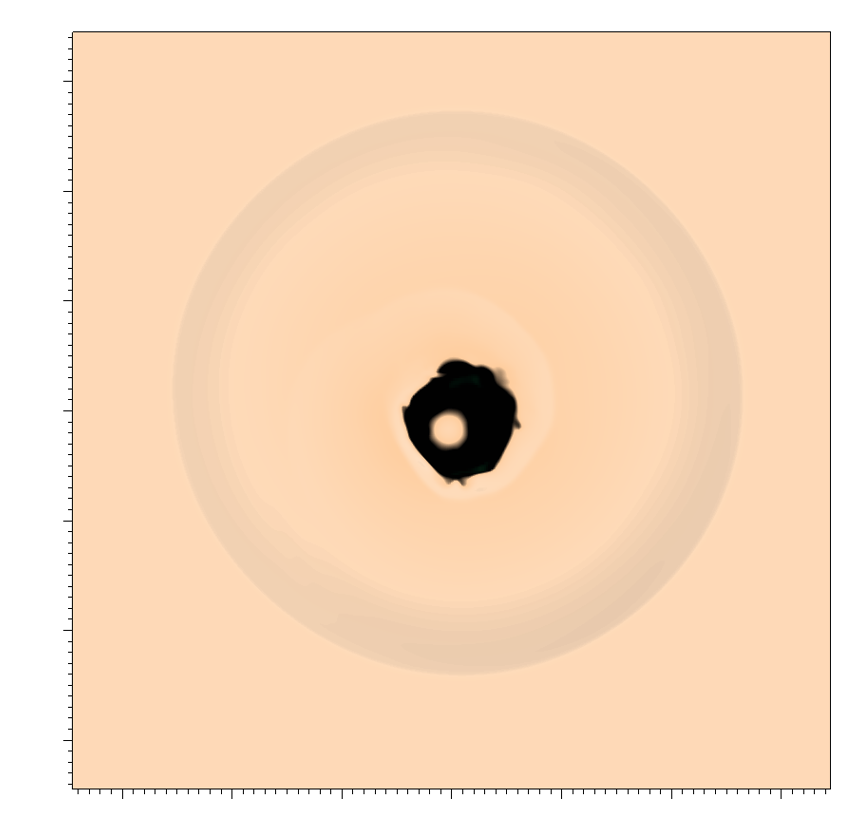}}
\subfloat[density and composition, xz plane]{\includegraphics[width=0.24\textwidth]{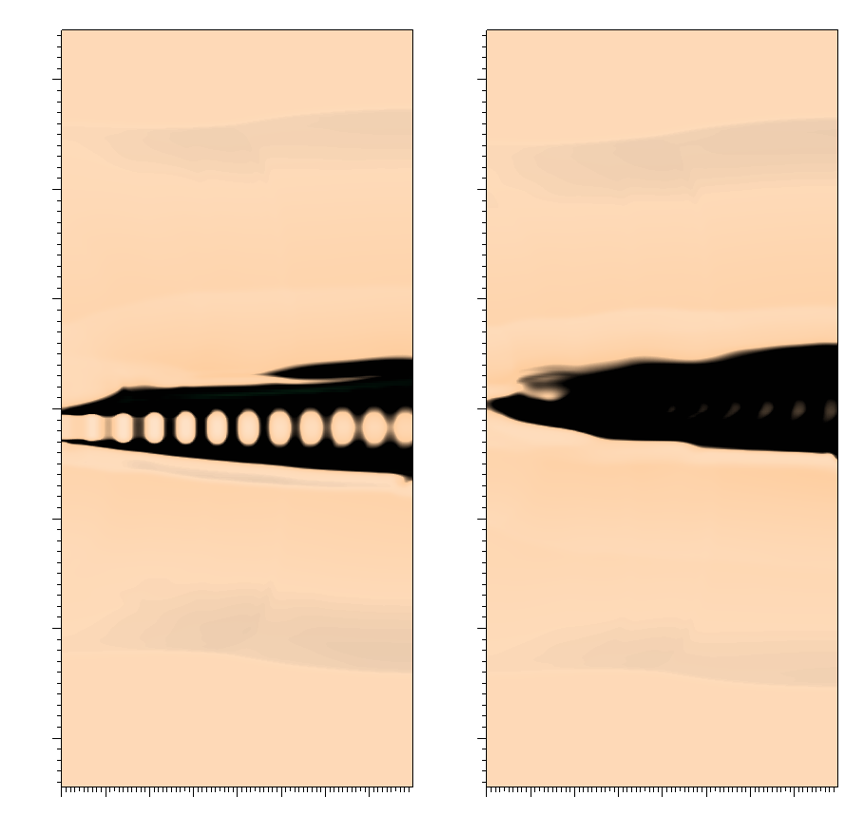}}\\
\subfloat[p, T and v at x=0\,cm]{\includegraphics[width=0.24\textwidth]{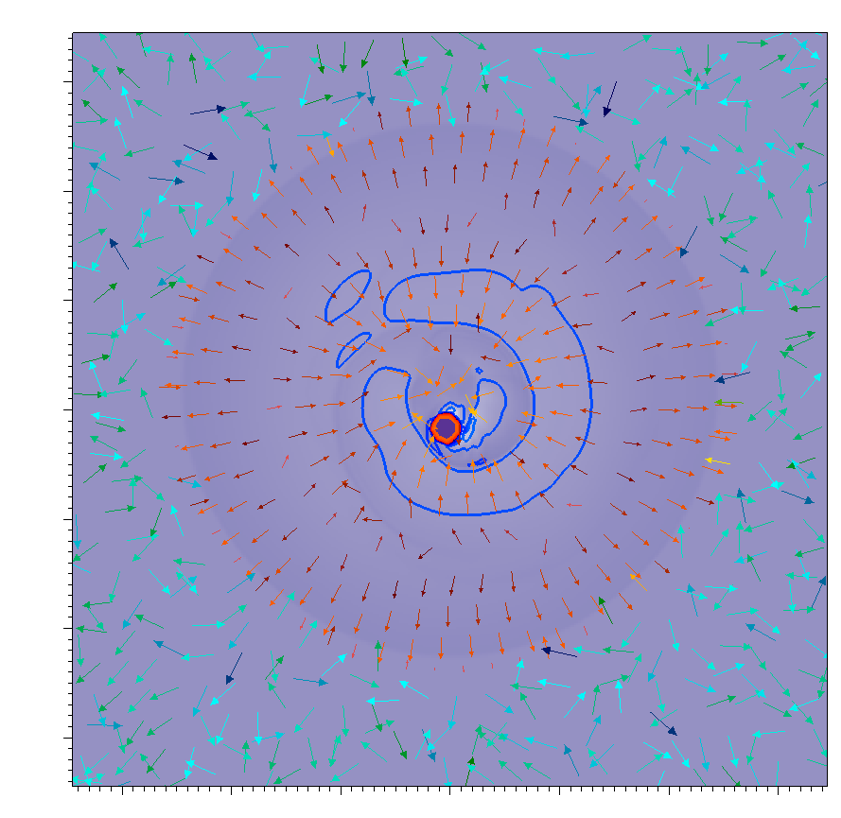}}
\subfloat[p, T and v at x=50AU]{\includegraphics[width=0.24\textwidth]{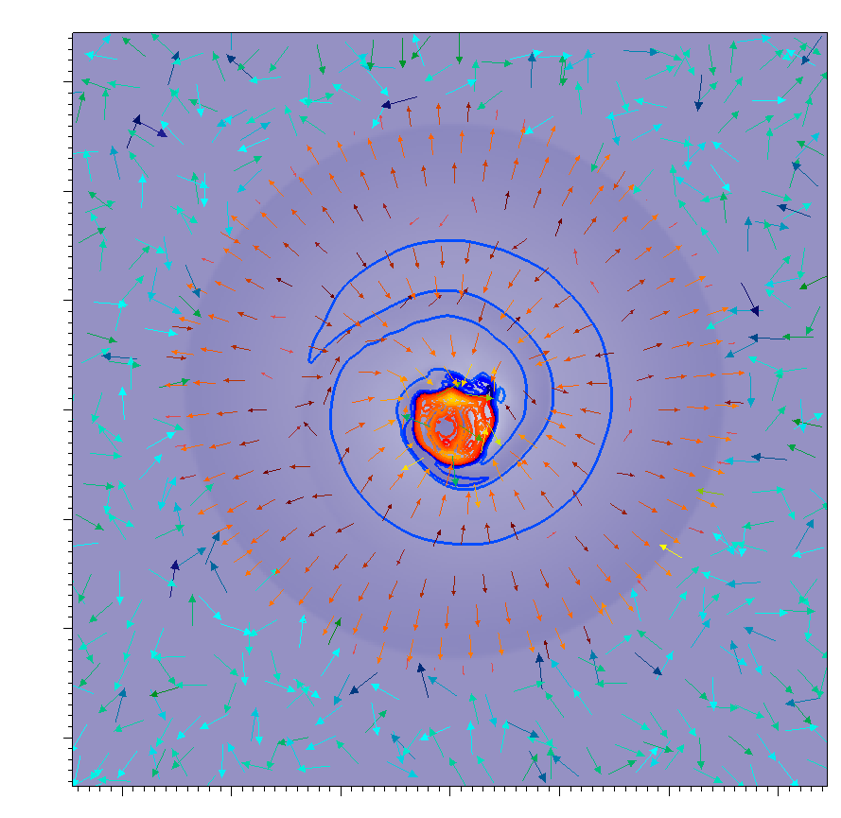}}
\subfloat[p, T and v at x=100AU]{\includegraphics[width=0.24\textwidth]{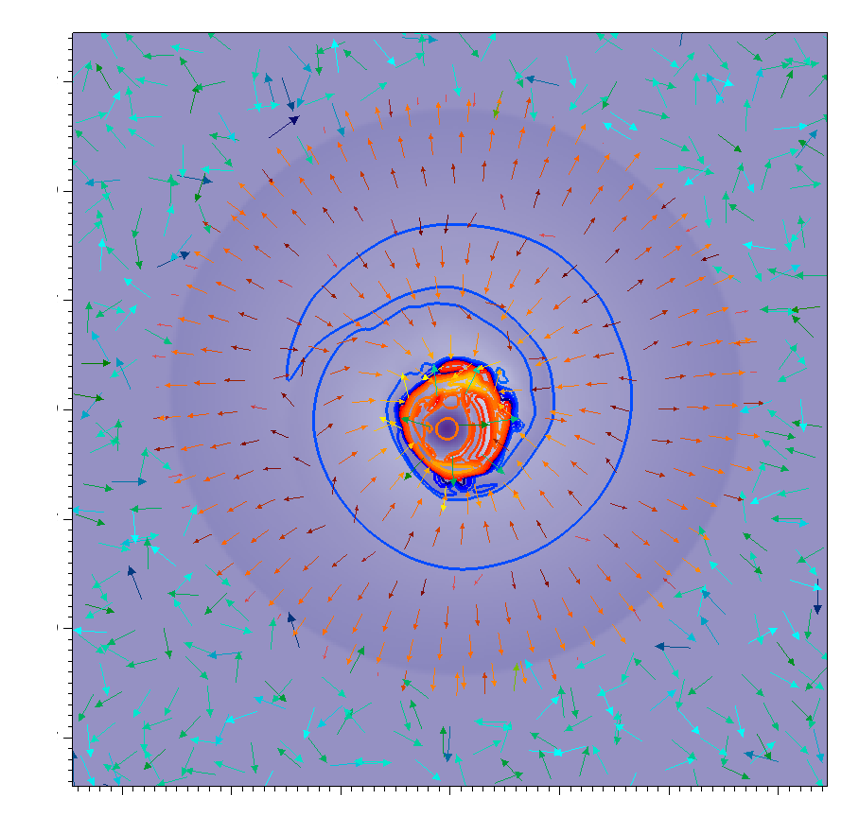}}
\subfloat[p, T and v in xy and xz planes]{\includegraphics[width=0.24\textwidth]{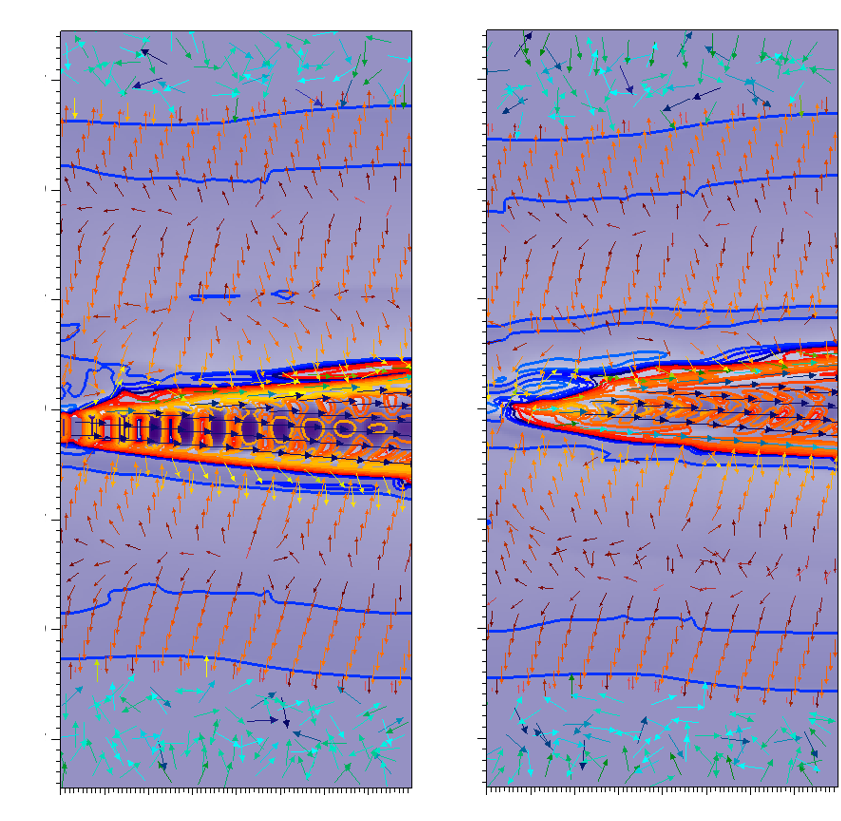}}
\\
\subfloat[density, H]{\includegraphics[width=0.14\linewidth]{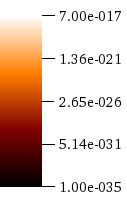}}
\subfloat[density, H$_2$]{\includegraphics[width=0.14\linewidth]{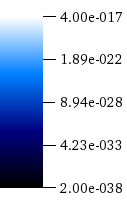}}
\subfloat[ion fraction]{\includegraphics[width=0.14\linewidth]{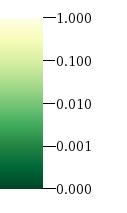}}
\subfloat[pressure]{\includegraphics[width=0.14\linewidth]{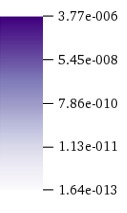}}
\subfloat[velocity field]{\includegraphics[width=0.14\linewidth]{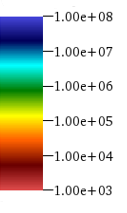}}
\subfloat[temperature]{\includegraphics[width=0.14\linewidth]{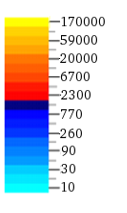}}\\
\caption[Co-Orbital Model s066.2.3: Section Plots, 175 years]{\textbf{Cross-sections of physical parameters for the single atomic jet in the co-orbital configuration}, s066.2.4: 18\,AU binary,  at simulation time 175 years.  Axis scales are in units of 10$^{15}$cm.  Ambient medium is atomic hydrogen with trace molecular hydrogen formed during the simulation.  Underlying density plot is fully opaque. }
\label{single-pressure}
\end{figure*}

Pressure, temperature and velocity fields of  the 18\,AU single atomic outflow at the simulation time of 175 years 
  are shown in Fig.\,\ref{single-pressure}.  Note that a scale for the partial density of H$_2$ appears on the plots since  trace H$_2$ appears, arising from the molecular cooling routine which models dust grain catalytic formation of molecules.   From these plots it is evident that the low-density cocoon surrounding the jet column, and also the low-density regions within the jet, are populated by hot, partially ionised material. 

The outer expanding shock is evident in the density distribution as a ring and in the pressure distribution as  a filled circle.  The shock introduces turbulence consistent with the  simulations of atomic winds with strong radiative cooling \citep[e.g.][]{2018MNRAS.480...75N}.
Examination of early-stage outputs in the simulation reveals that material driven outwards by the expanding outflows passes through this shock and then becomes disorganised.  This can be seen occurring in the time-stepped images of the dual atomic-molecular outflow version of this model.

A common feature to all the models is lateral flow expansion.  This is expected as all of the outflows (atomic and molecular) are over-pressured with 
respect to the ambient medium (see Table \ref{tab:commonparams}).  Based on an estimated circular radius computed from the jet cross-sectional area (see Fig.\ref{6623_jetbend_misc}) an approximate half-opening angle for the jet of the s066.2.4 simulation is 8.5$^{\circ}$ degrees.

As the expanding material pushes outwards in its forward progress, it  transfers some of its  momentum in the x-direction to the 
ambient material in the boundary layer. This creates an entrained updraft.  This is evident in all the 
simulations from the region of lower pressure surrounding the jet columns, and the surrounding velocity fields.  There is no evidence of Kelvin-Helmholtz mixing of medium and jet material in the boundary layer.  It should be noted that the prototyping long jet simulations did not indicate such features until after 200\,AU of jet propagation (see Fig.\,\ref{proto1}).

When the main outflow bow shock has left the problem domain, the return flow (backflow) is not adequately 
modelled.  This phenomenon, a potential inaccurate feature of simulations,  is discussed in \citet{1982A&A...113..285N}.  However, the 
prototyping phase did not show evidence of any significant backflow.   It was decided that backflow was not a feature that was likely to have a substantial influence on the dynamics of our jets in the region of interest here.  This is particularly true of our main simulations which include the 
dense molecular outflow which will have a greater effect on the atomic outflow than a relatively weak backflow.  Also of greater impact is the 
orbital motion of the inlets which will perturb the jet column to a greater extent than vortex shedding from the bow shock, which has its main 
impact close to the head of the jet column.  Based on the prototype models, the head of the jet is 800\,AU+ distant during the simulation timespan of 
our key results.

\subsection{Co-orbital Model s066.2.3: Time Evolution}

\begin{figure}
\subfloat[z-y plane, $x = 0$cm]{\includegraphics[width=0.49\linewidth]{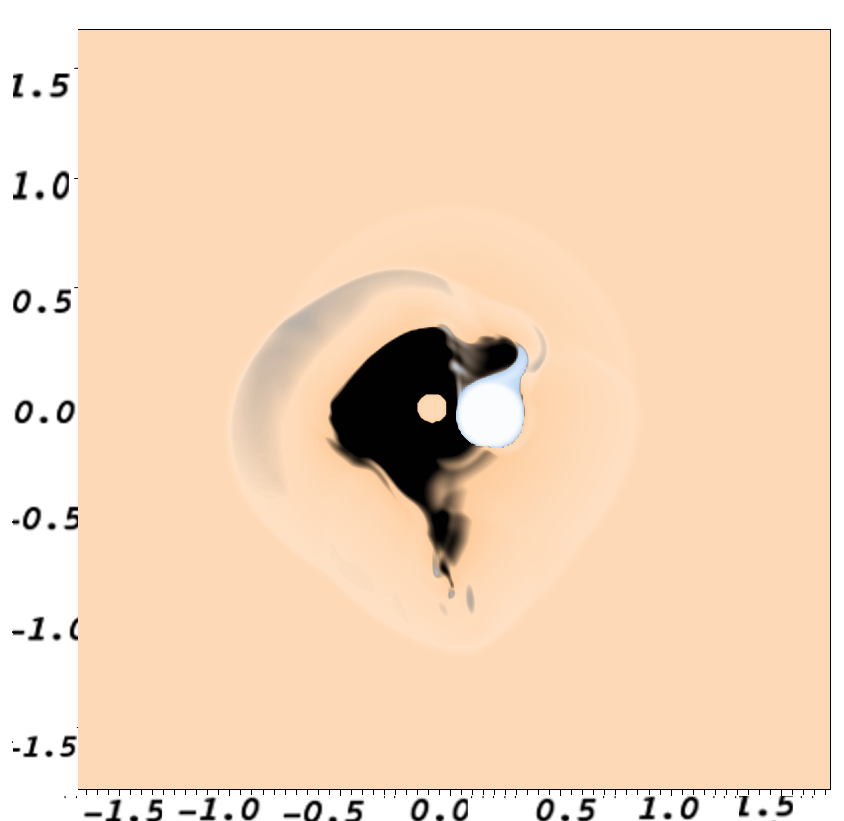}} 
\subfloat[z-y plane, $x = 7.5 \times 10^{14}$ cm]{\includegraphics[width=0.49\linewidth]{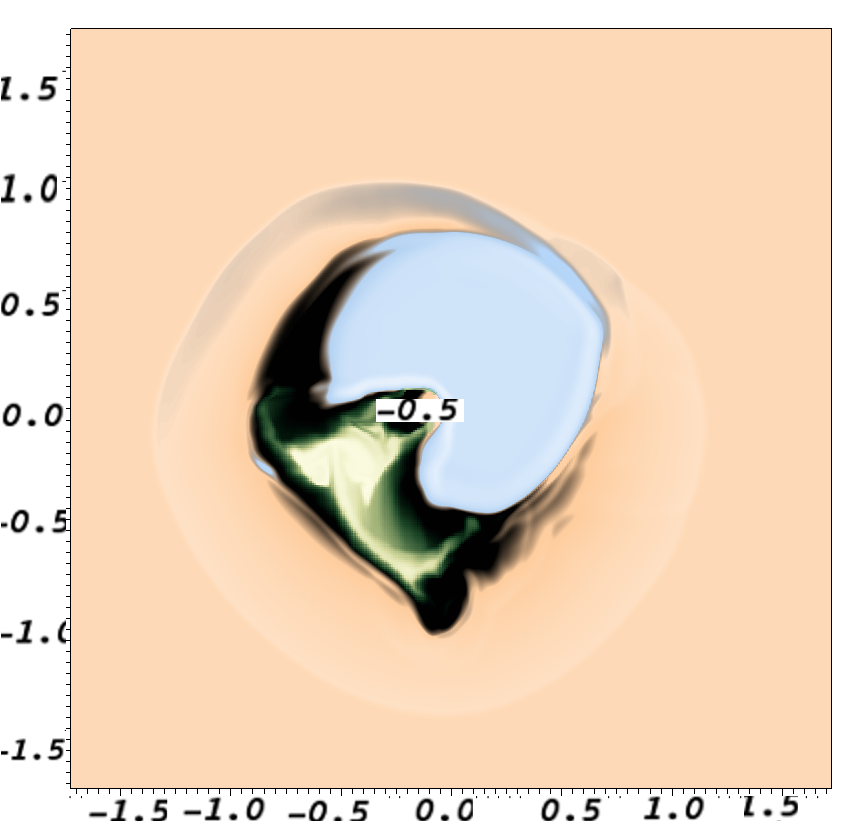}} \\\\
\subfloat[z-y plane, $x = 1.5 \times 10^{15}$ cm]{\includegraphics[width=0.49\linewidth]{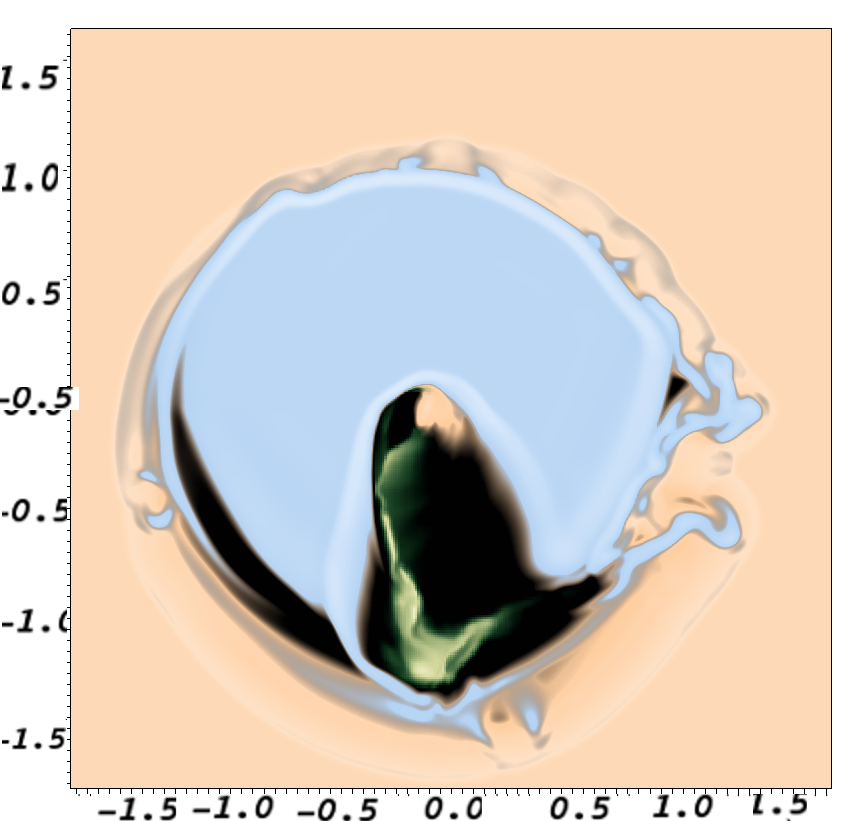}}
\subfloat[x-y plane, z = 0cm; $x-z$ plane, $y = 0$cm]{\includegraphics[width=0.49\linewidth]{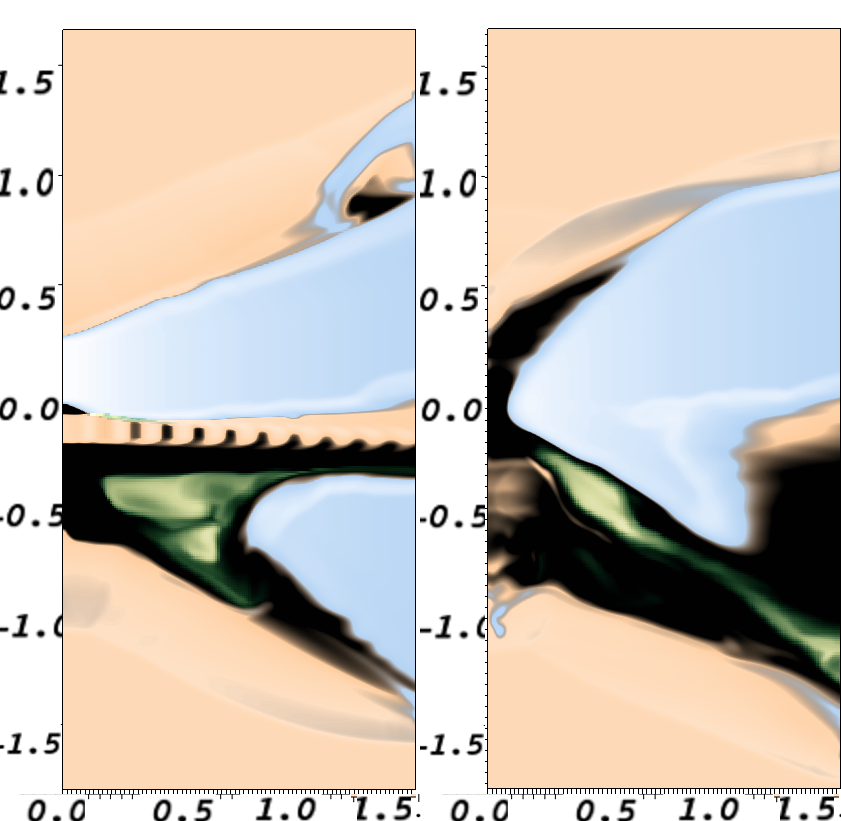}}\\\\
\hspace*{0.6cm} \subfloat[log(density, gcm$^{-3})$]{\includegraphics[width=0.30\linewidth]{./Images/6624_1300_legend_density.png}}
\subfloat[log(H$_2$ density, gcm$^{-3}$)]{\includegraphics[width=0.30\linewidth]{./Images/6624_1300_legend_H2.png}}
\subfloat[log(ion fraction)]{\includegraphics[width=0.30\linewidth]{./Images/6624_1300_legend_ion.png}}
\\
\caption[Co-Orbital Model s066.2.3: Section Plots, 87.5 years]{\textbf{Co-Orbital Model s066.2.3: 18AU Binary, atomic-molecular outflow, density  cross-sections at simulation time $t=87.5$~years.}  Axis scales are in units of 10$^{15}$cm.  Ambient medium is atomic hydrogen with trace molecular hydrogen formed from cooling during the simulation.  Underlying density plot is fully opaque.  For clarity, H$_2$ and ionisation fraction overplots are at ramped opacity; 100\% opacity at max value, transparent at minimum value.}
\label{6623_1300_xsect}
\end{figure}

The molecular outflow, emerging from the secondary binary partner, is now introduced.
 Figure \ref{6623_1300_xsect} shows a set of density  cross-sectional  plots at the simulation time  87.5 years.  As before, this figure gives an overview of the outflow structure,  sharing a common colour scaling of variables.  

As with the single-jet orbital simulation, a spiral density wave radiates outwards from the jet column, induced in the surrounding medium by the orbital motion of the jet.  However, when we examine the dual outflow system at 100\,AU along the x-axis (see panel (c) of Figure \ref{6623_1300_xsect}) we find that the expanding molecular flow is overtaking the density wave.   An approximate calculation finds that the molecular flow 
expands at $\sim$ 6 km\,s$^{-1}$ radially from the x-axis as it propagates 
 across the domain; whilst the sound speed in the ambient medium is 1.09 km\,s$^{-1}$ (see Table \ref{tab:commonparams}).  

\begin{figure*}
\subfloat[density and composition, x=0\,AU]{\includegraphics[width=0.24\textwidth]{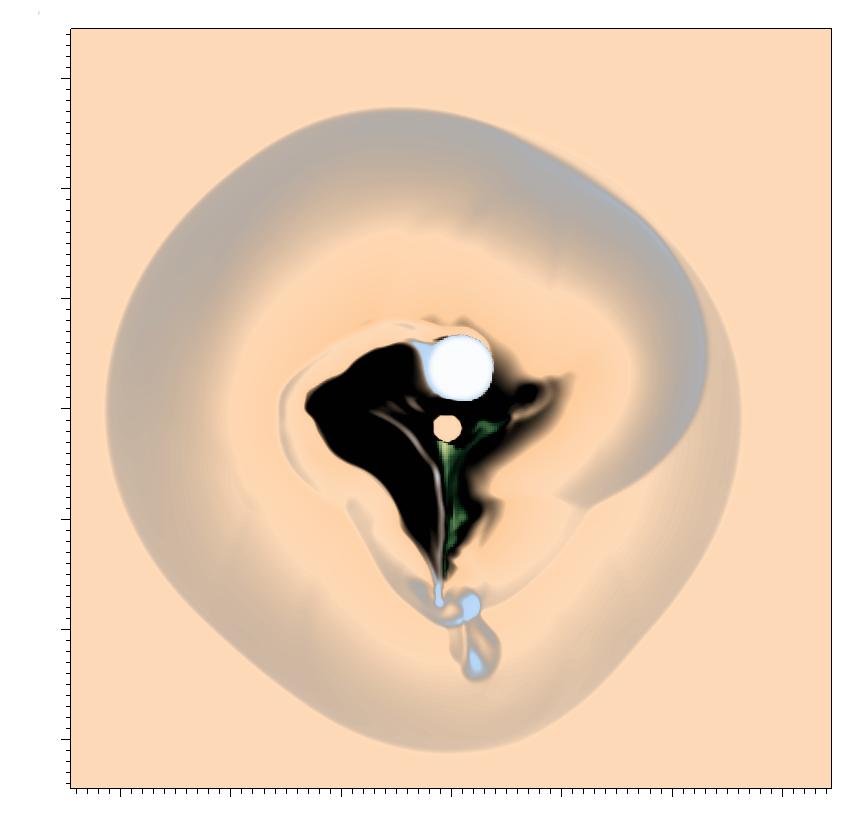}}
\subfloat[density and composition, 50AU]{\includegraphics[width=0.24\textwidth]{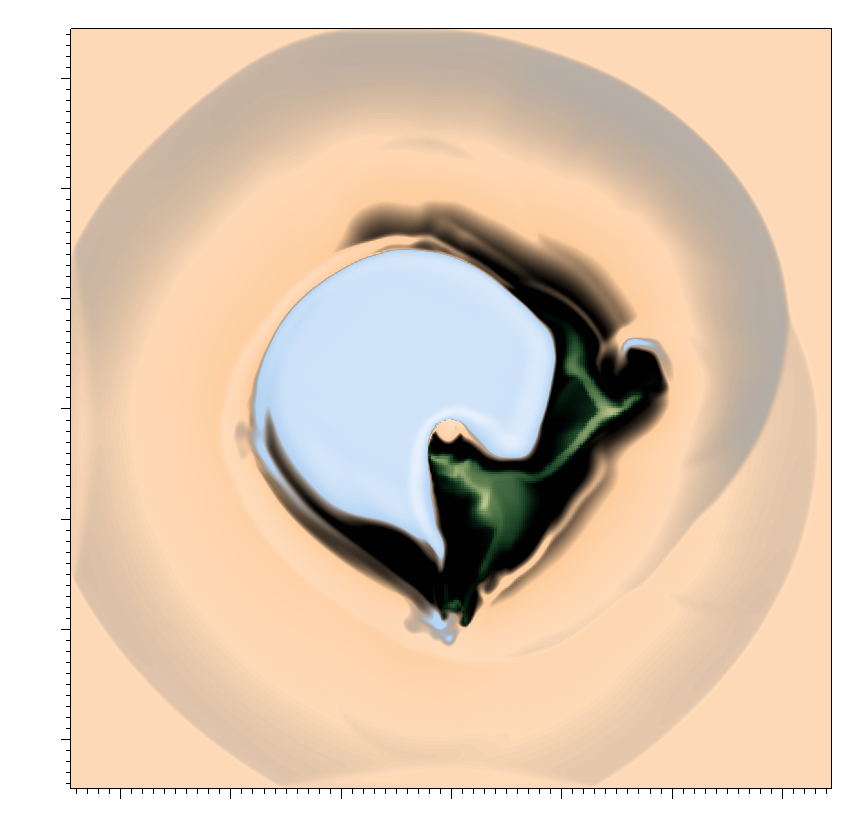}}
\subfloat[density and composition, 100AU]{\includegraphics[width=0.24\textwidth]{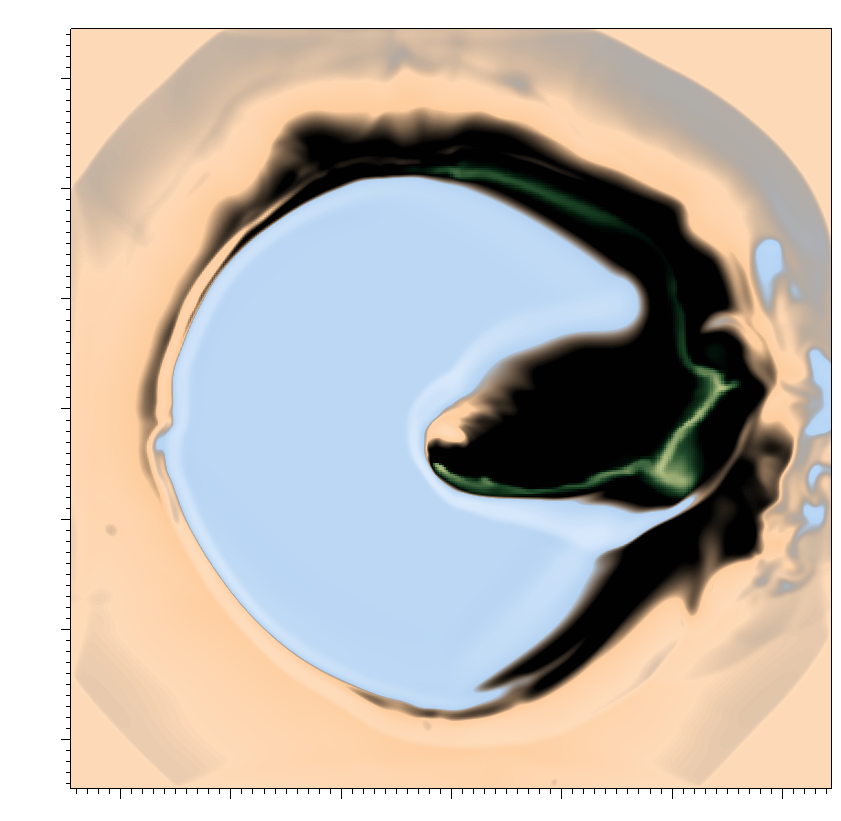}}
\subfloat[density, composition, xy, xz plane]{\includegraphics[width=0.24\textwidth]{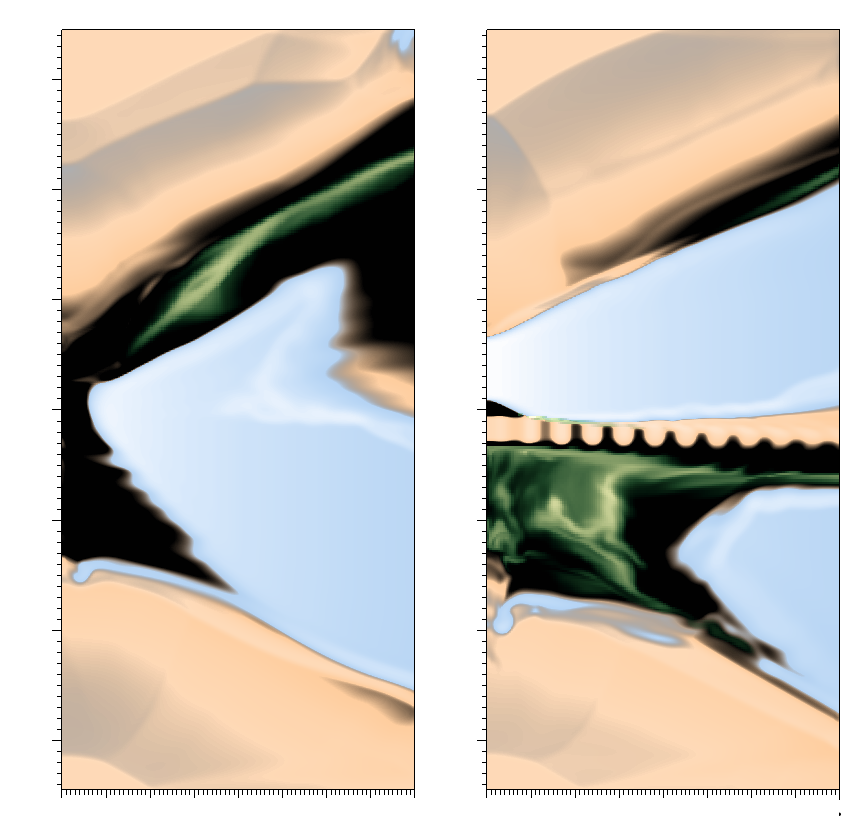}}\\
\subfloat[p, T and v at x=0\,AU]{\includegraphics[width=0.24\textwidth]{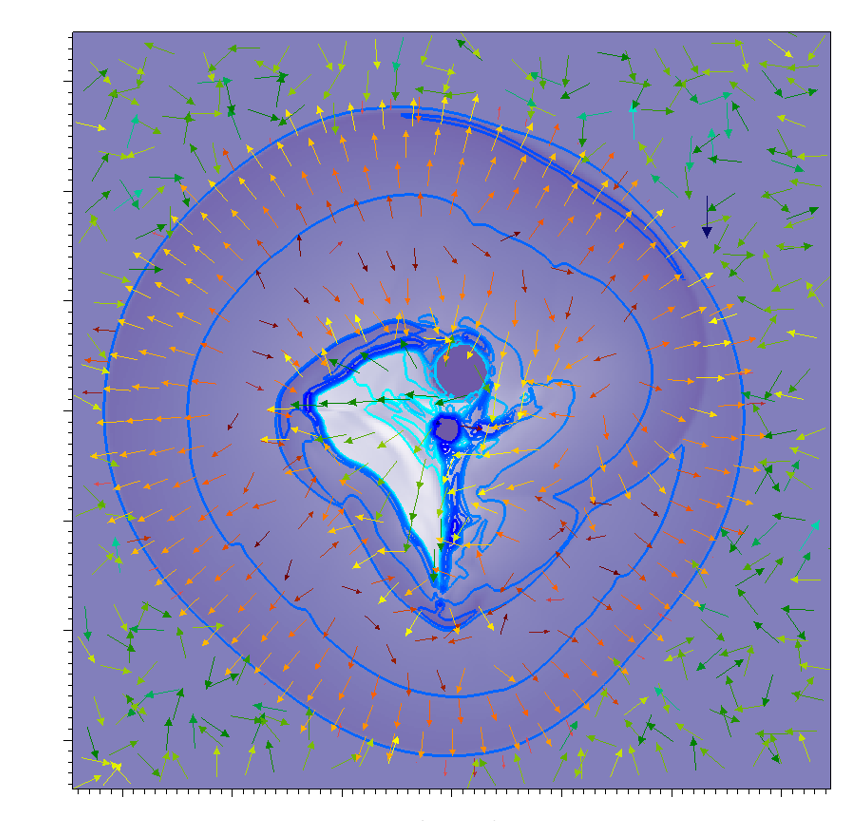}}
\subfloat[p, T and v at x=50AU]{\includegraphics[width=0.24\textwidth]{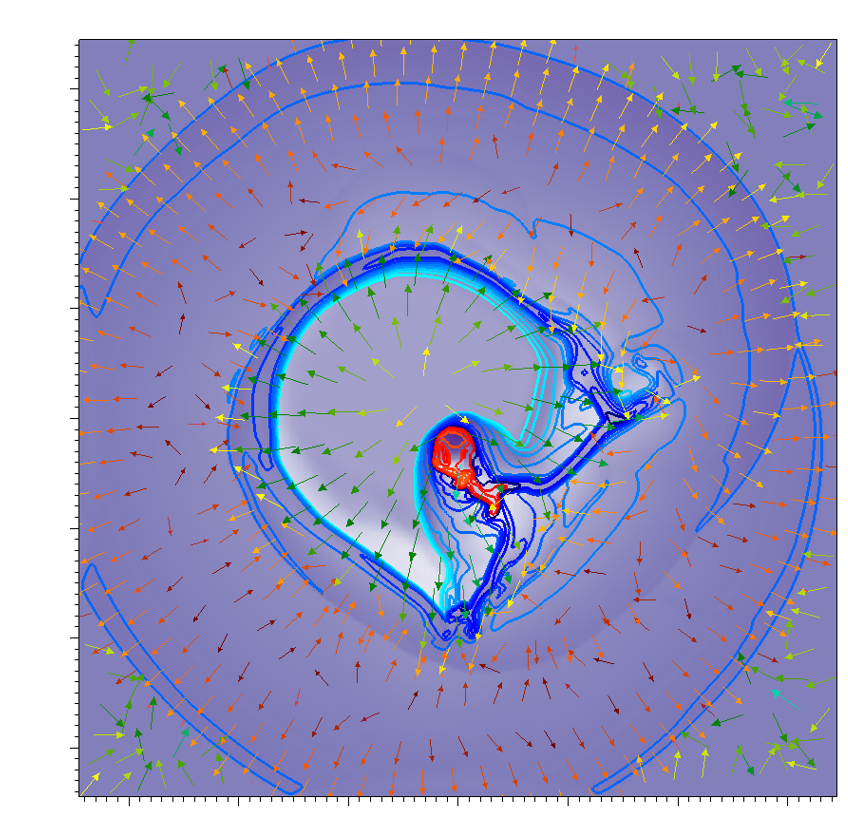}}
\subfloat[p, T and v at x=100AU]{\includegraphics[width=0.24\textwidth]{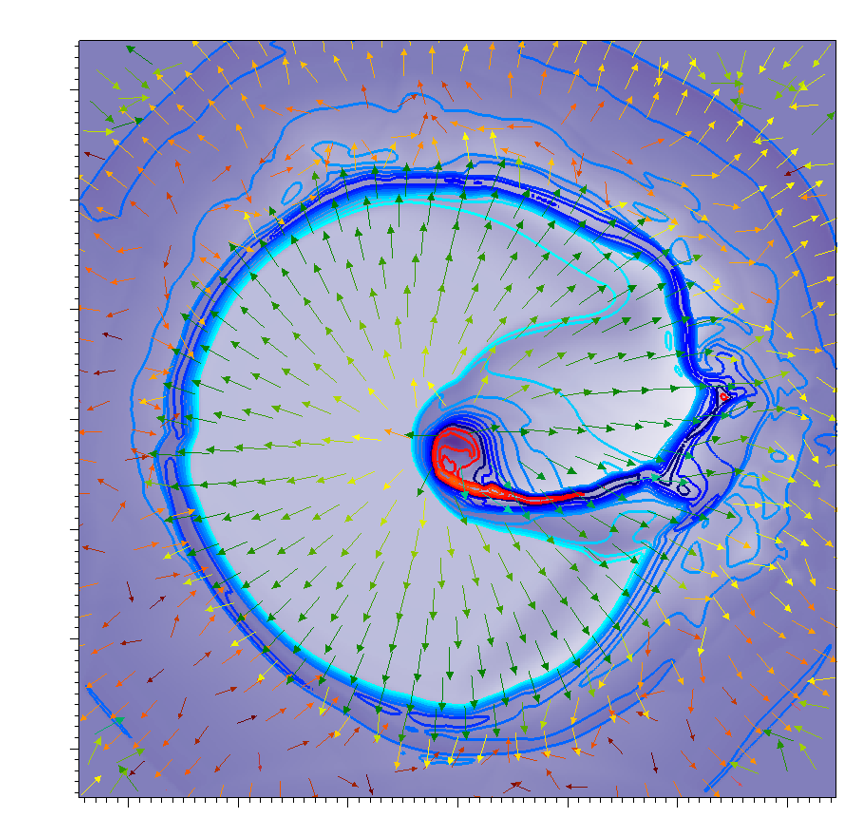}}
\subfloat[p, T and v in xy and xz planes]{\includegraphics[width=0.24\textwidth]{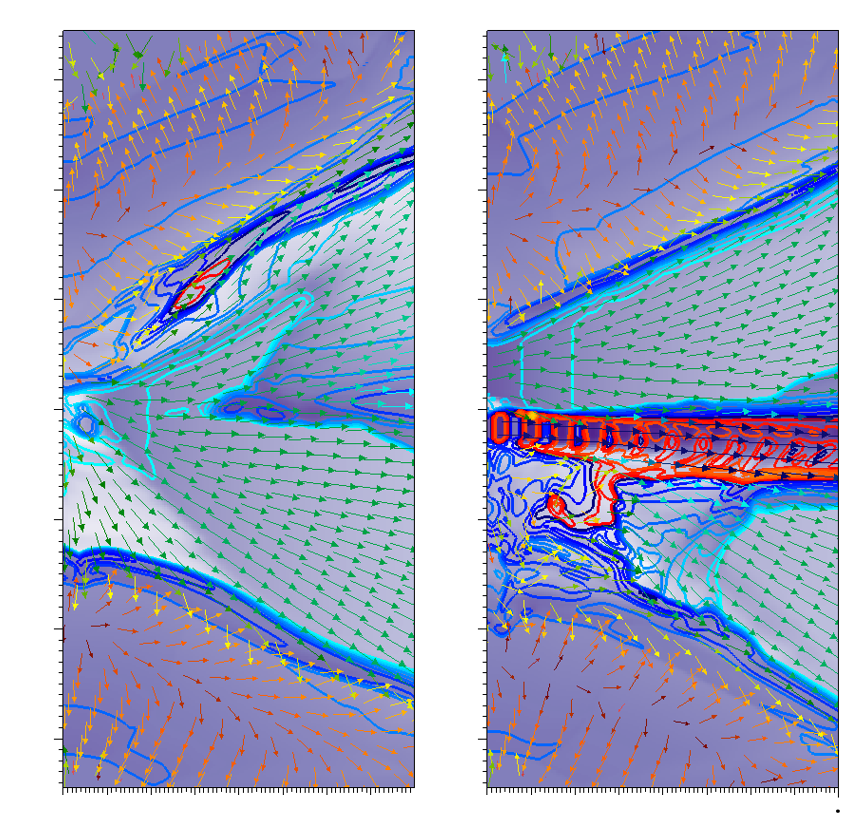}}
\\
\subfloat[density, H]{\includegraphics[width=0.14\linewidth]{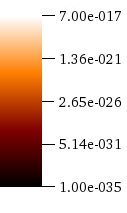}}
\subfloat[density, H$_2$]{\includegraphics[width=0.14\linewidth]{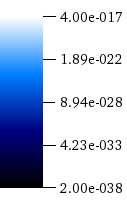}}
\subfloat[ion fraction]{\includegraphics[width=0.14\linewidth]{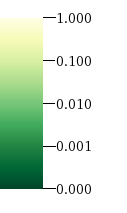}}
\subfloat[pressure]{\includegraphics[width=0.14\linewidth]{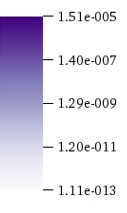}}
\subfloat[velocity field]{\includegraphics[width=0.14\linewidth]{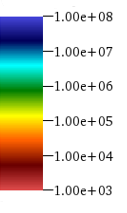}}
\subfloat[temperature]{\includegraphics[width=0.14\linewidth]{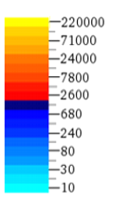}}\\
\caption[Co-Orbital Model s066.2.3: Section Plots, 175 years]{\textbf{Cross-sections of physical parameters for the Co-Orbital Model} s066.2.3: 18\,AU Binary, atomic-molecular outflow at simulation time 175 years.  Axis scales are in units of 10$^{15}$cm.  Ambient medium is atomic hydrogen with trace molecular hydrogen formed during the simulation.  Underlying density plot is fully opaque. }
\label{6623_2599}
\end{figure*}

In the early stages of propagation the molecular outflow is feeding material into the spiral density wave, which carries this material outwards just behind its advancing shock.  This can be seen much more vividly in Figure \ref{6623_2599}, which shows the flow at a more fully developed stage at 175 years.  By this stage, the slower moving shock has caught up with and overtaken the molecular ouflow at x = 100\,AU, since the opening angle of the molecular flow remains unchanged between this and the 87.5 year stage in Fig.\ref{6623_1300_xsect}.  

In addition to a well-mixed partial density of H$_2$ molecules in the expanding spiral  wave, there are gobbets of wholly molecular material.  In Panel (c) of Fig.\,\ref{6623_1300_xsect}, we find evidence of shearing instability pulling molecular material out from the main outflow where it interacts with the shock. 

 In Figure \ref{6623_2599}  it appears that lumps of H$_2$ are being flung outwards, possibly by the action of the atomic and molecular outflow columns from co-orbiting sources, as they stir the surrounding medium like a giant cosmic egg-whisk.   
 
We see in Figure \ref{6623_2599}, panels (d) and (h), that the pulsed velocity signal has given rise to small-scale density knots within the jet column, sandwiched between regions of lower density, just as we saw in the atomic-only case.  However from around x = 20\,AU onwards, the atomic jet is in direct contact with the molecular outflow along an advancing face, and the integrity of the knots is severely compromised, as a large crossing shock from the impact point sweeps through the jet while the flow carries it forwards.

There is substantial ionisation occurring in this model, indicated by a green-white colour scale in the plots.  The ionised material can be seen mainly in regions of very low density.  However a more detailed examination will show that these are not the regions where ionisation is occurring.
 Panels (c) and (g)  of Fig.~\ref{6623_2599} demonstrate that the atomic outflow does not pass peacefully through its wide-angle molecular companion.  A large `hole' has been blasted out of the side of the conical molecular flow.  This  gives a clue as to the origin of the ionised material: it is pouring out of the region where the leading face of the atomic jet is in contact with, and ploughing through, the molecular outflow.

Examination of Panel (e) of Fig. \ref{6623_2599} reveals an interesting feature. To the `South' and `West' of the barycentre of the two-jet system is a region of very low pressure ($\sim$ 10$^{-13}$ Pa) and temperature ($\sim$ 10\,K) which has formed.  The velocity field indicates that the inflation of this cavity and drop in pressure and temperature is due to a rapid expansion of material, most of which appears to be cast off from the molecular outflow column.

\begin{figure*}
\subfloat{\includegraphics[width=0.50\linewidth]{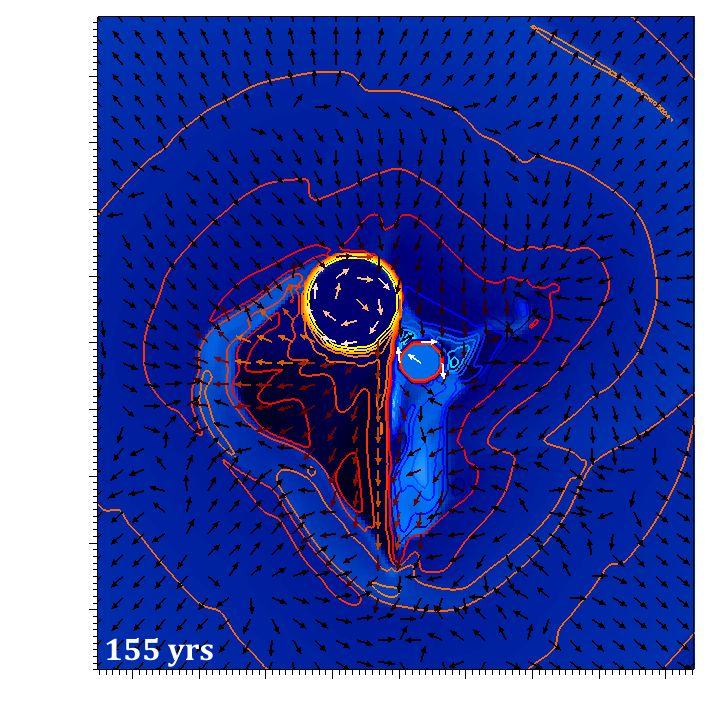}}
\subfloat{\includegraphics[width=0.49\linewidth]{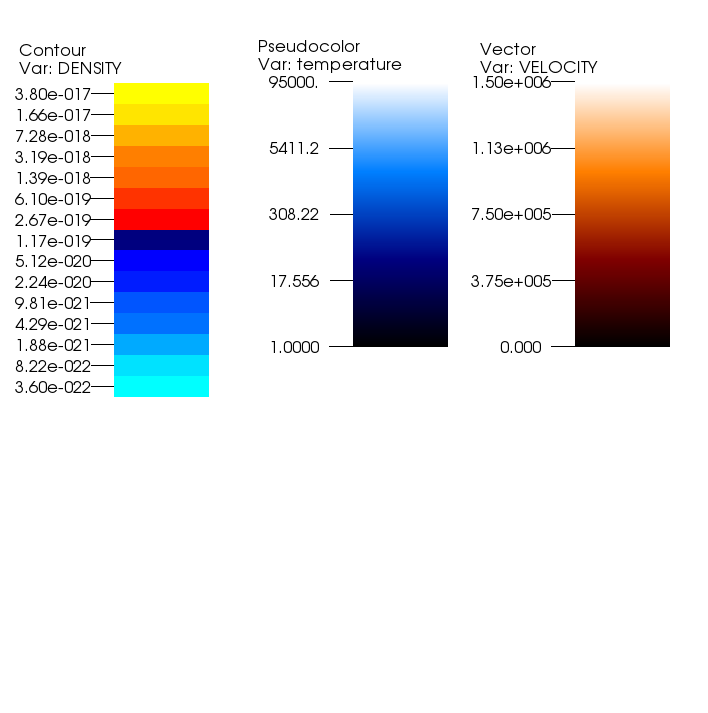}}\\[-2.5ex]
\caption[18AU Atomic Primary, Molecular Secondary, Dynamic Interaction]{Atomic Primary, Molecular Secondary, orbitally driven interaction; cross section at
155 years at x=3.2$\times$10$^{13}$ cm.  Background  (blue) shows temperature.  Velocity vectors are constant length, colour-scaled.  Contours show density.  Model number: s066.2.3}
\label{delavalone}
\end{figure*}

\begin{figure*}
\subfloat{\includegraphics[width=0.20\linewidth]{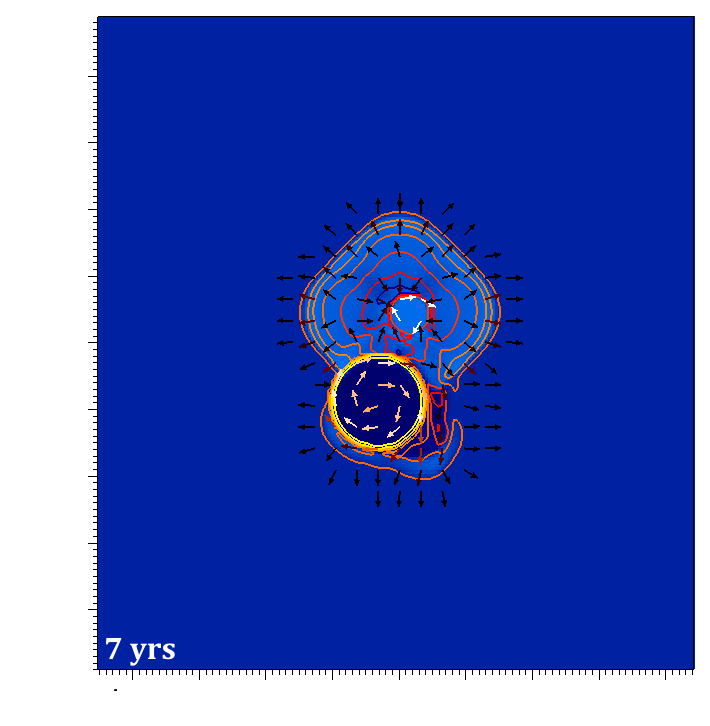}}
\subfloat{\includegraphics[width=0.20\linewidth]{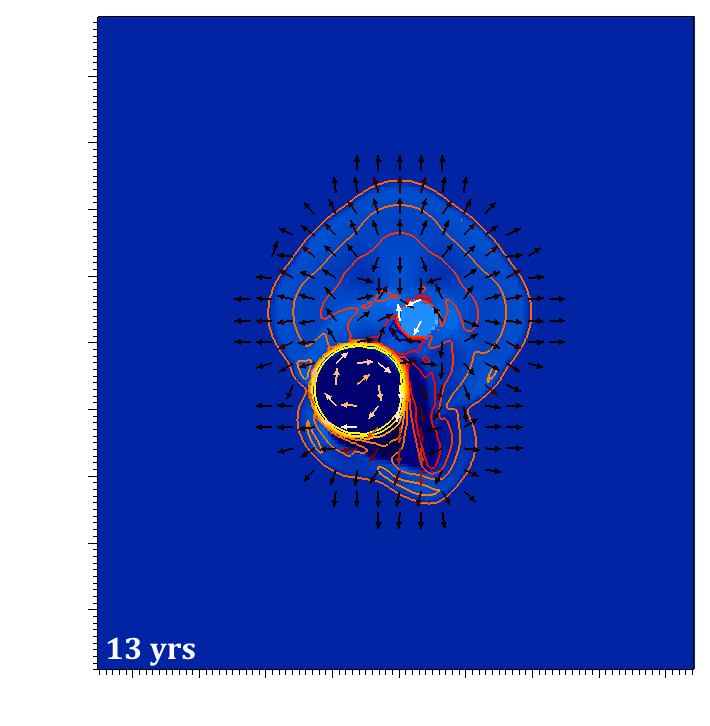}}
\subfloat{\includegraphics[width=0.20\linewidth]{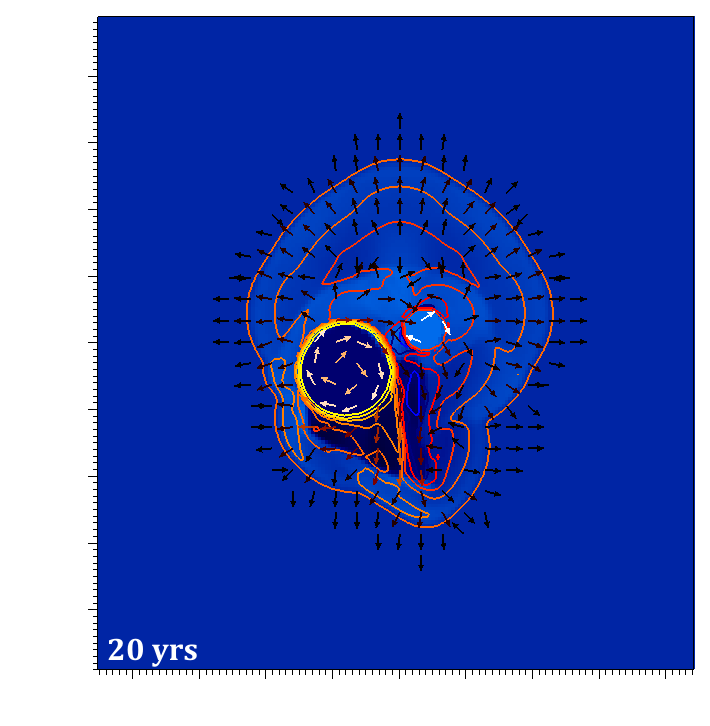}}
\subfloat{\includegraphics[width=0.20\linewidth]{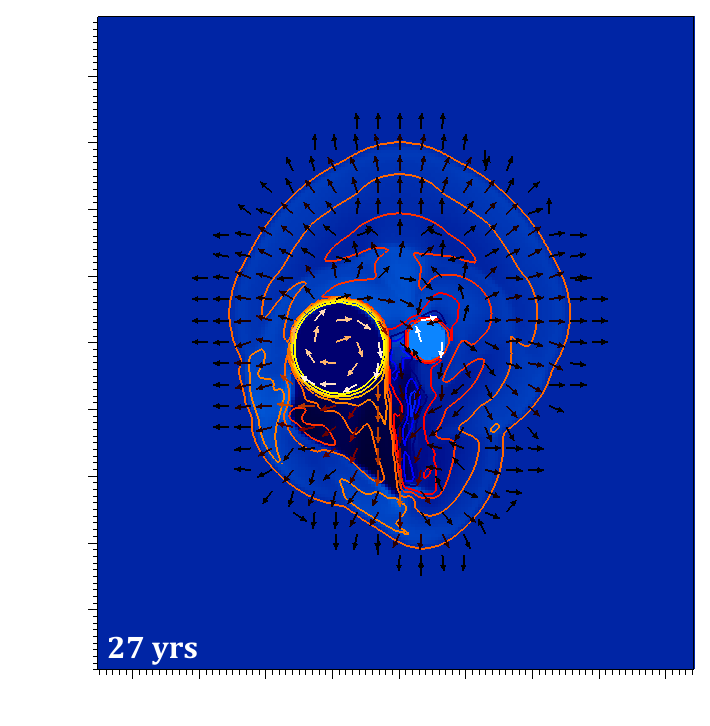}}\\[-2.5ex]
\subfloat{\includegraphics[width=0.20\linewidth]{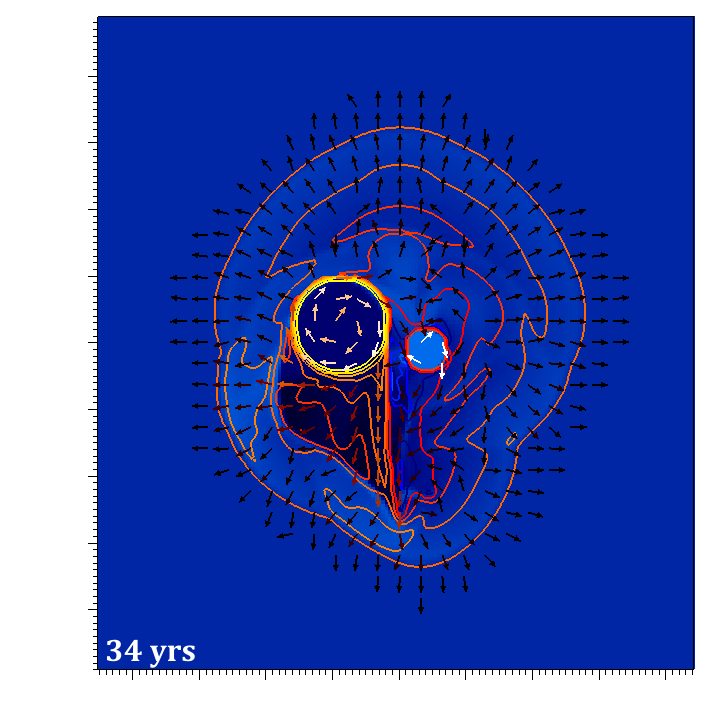}}
\subfloat{\includegraphics[width=0.20\linewidth]{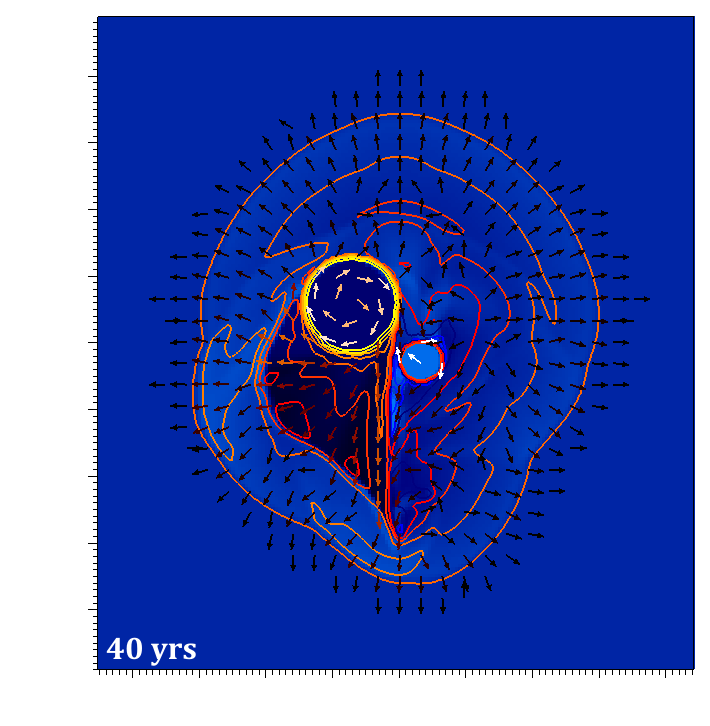}}
\subfloat{\includegraphics[width=0.20\linewidth]{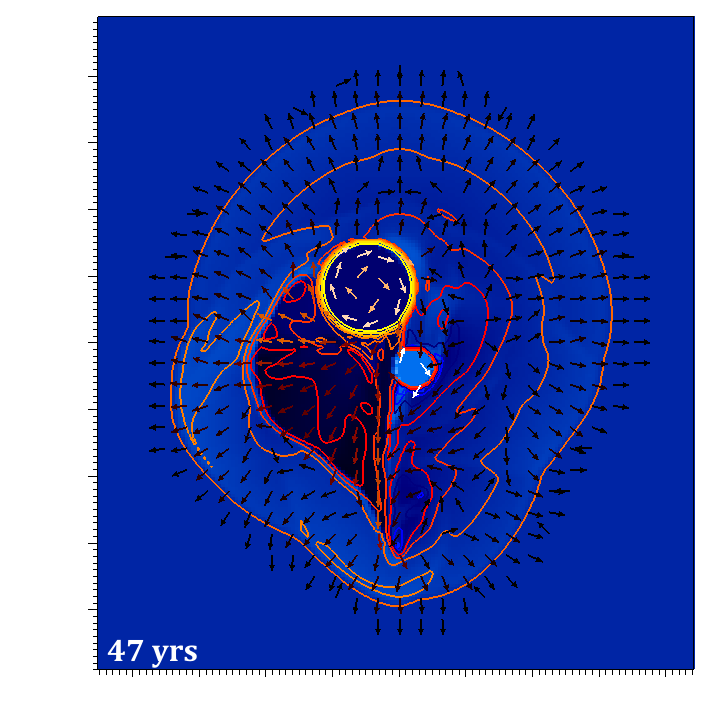}}
\subfloat{\includegraphics[width=0.20\linewidth]{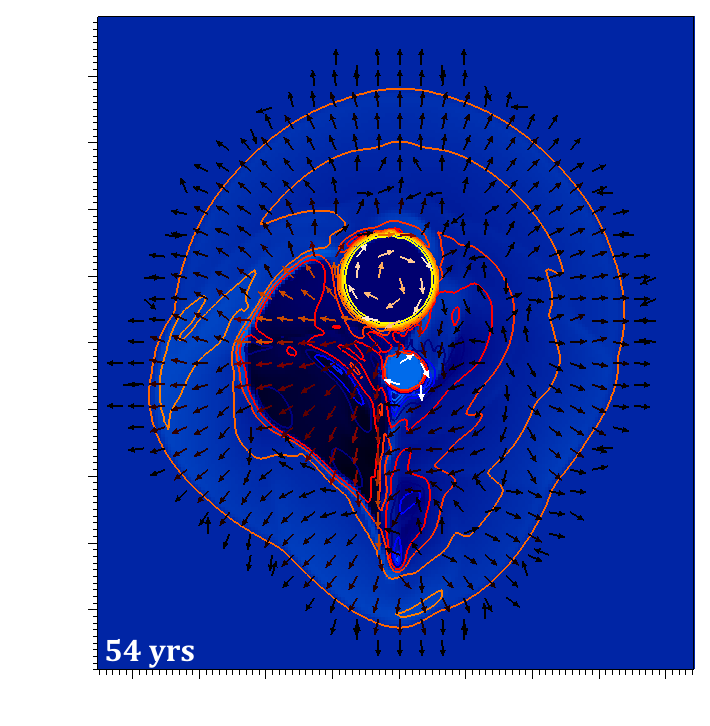}}\\[-2.5ex] 
\subfloat{\includegraphics[width=0.20\linewidth]{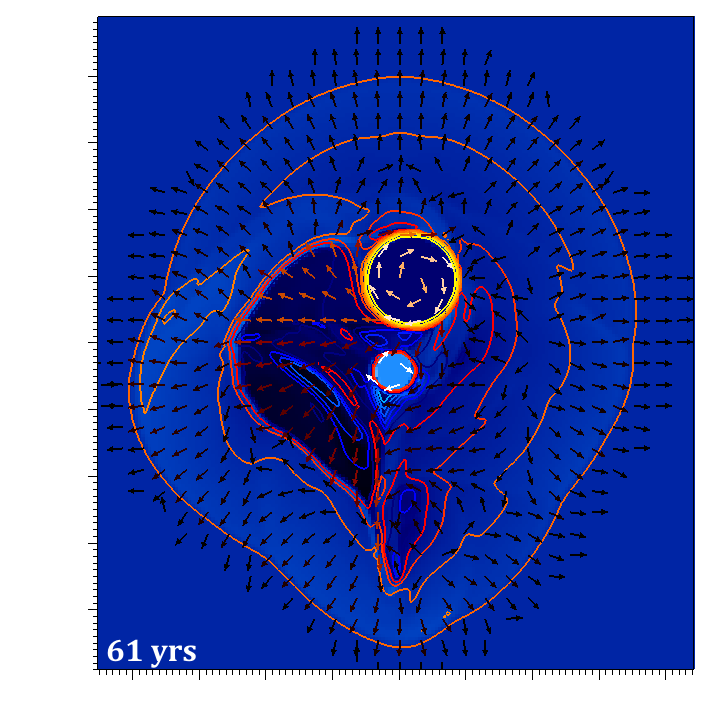}}
\subfloat{\includegraphics[width=0.20\linewidth]{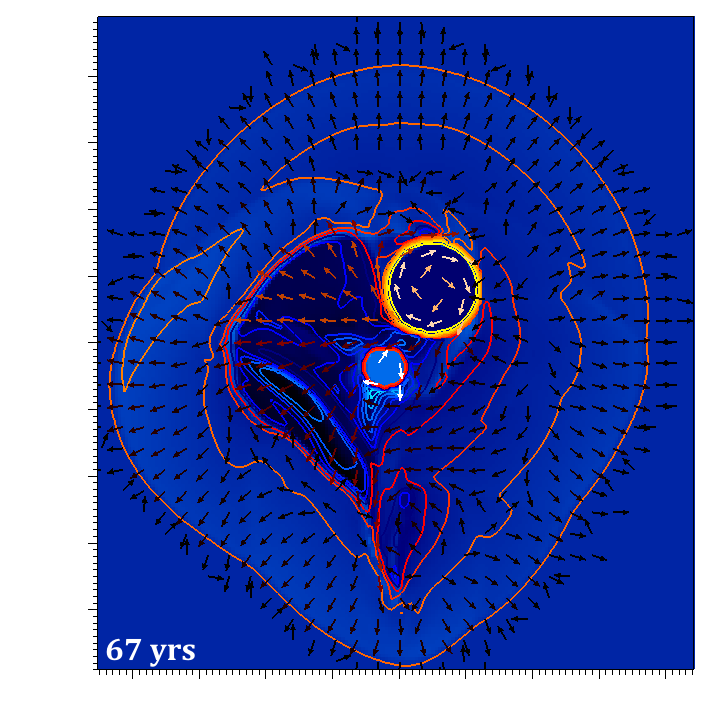}}
\subfloat{\includegraphics[width=0.20\linewidth]{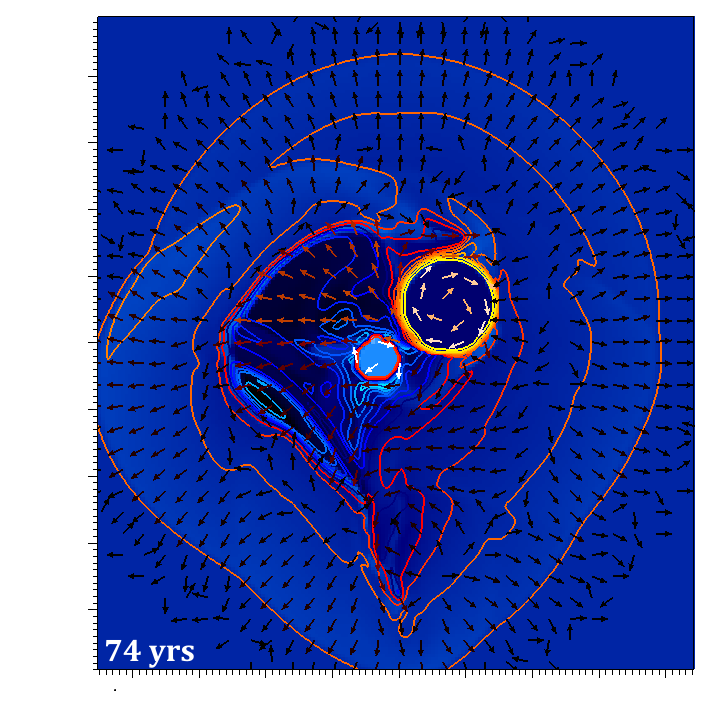}}
\subfloat{\includegraphics[width=0.20\linewidth]{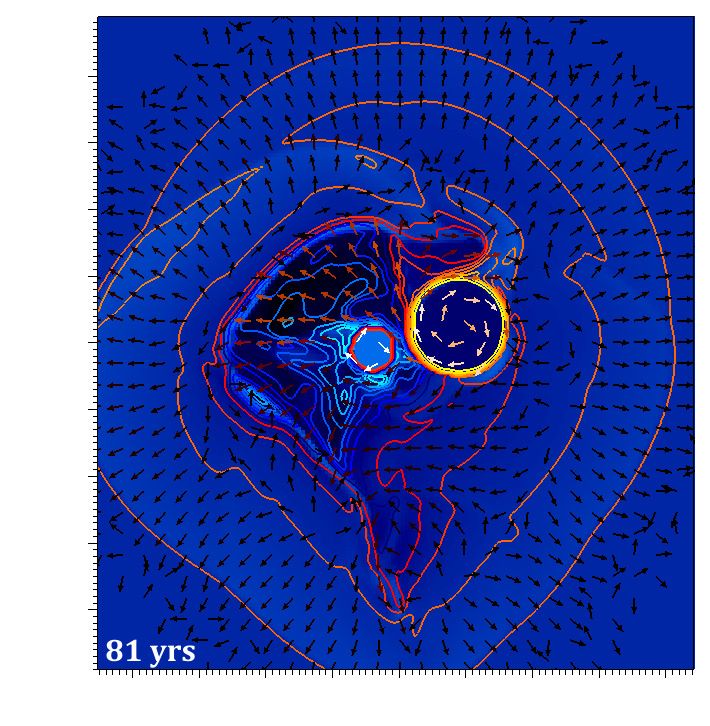}}\\[-2.5ex]
\subfloat{\includegraphics[width=0.20\linewidth]{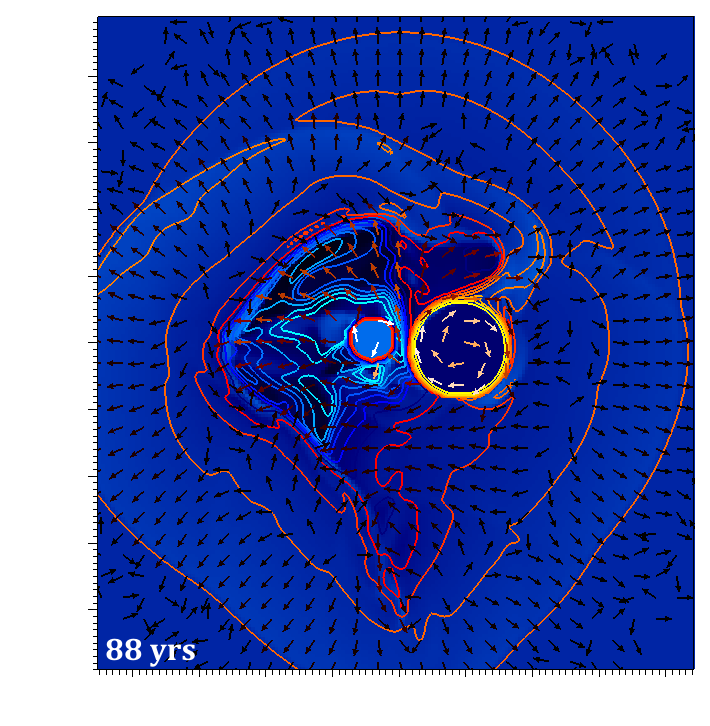}}
\subfloat{\includegraphics[width=0.20\linewidth]{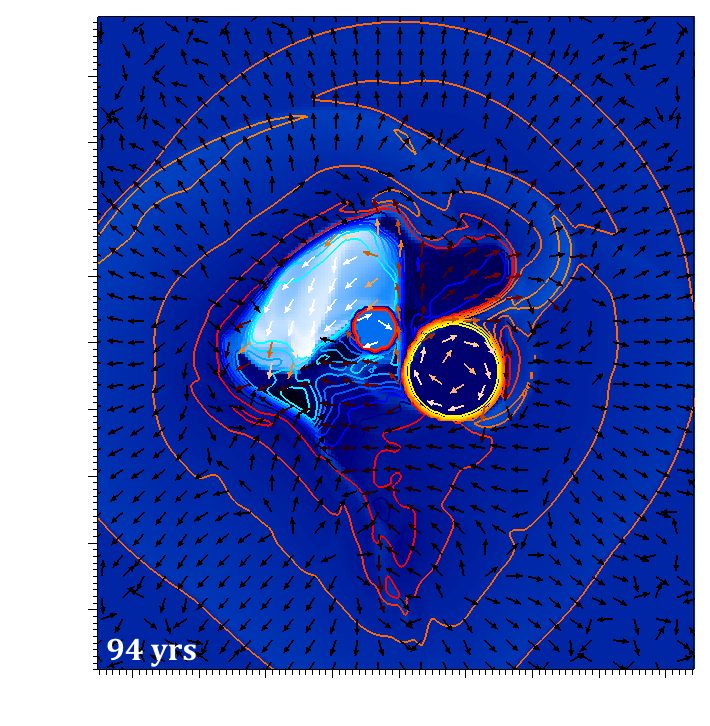}}
\subfloat{\includegraphics[width=0.20\linewidth]{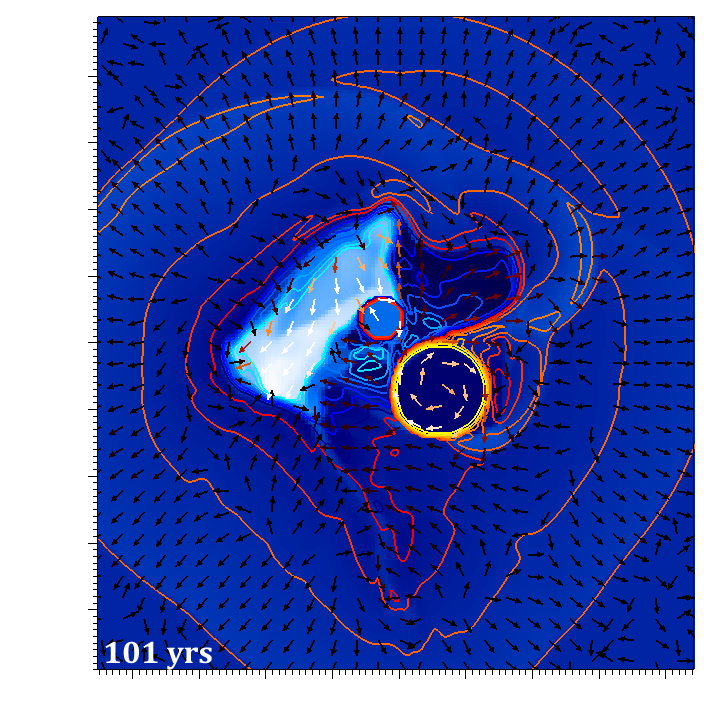}}
\subfloat{\includegraphics[width=0.20\linewidth]{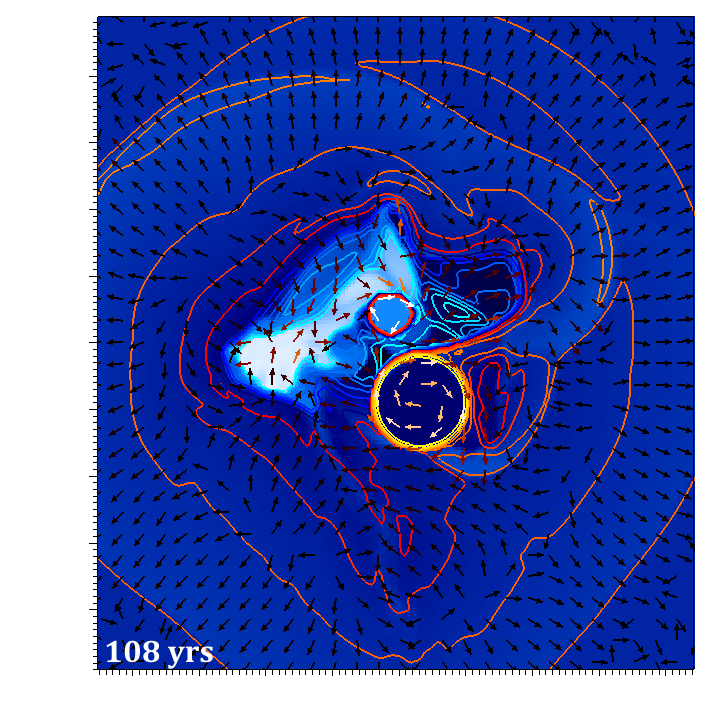}}\\[-2.5ex]
\subfloat{\includegraphics[width=0.20\linewidth]{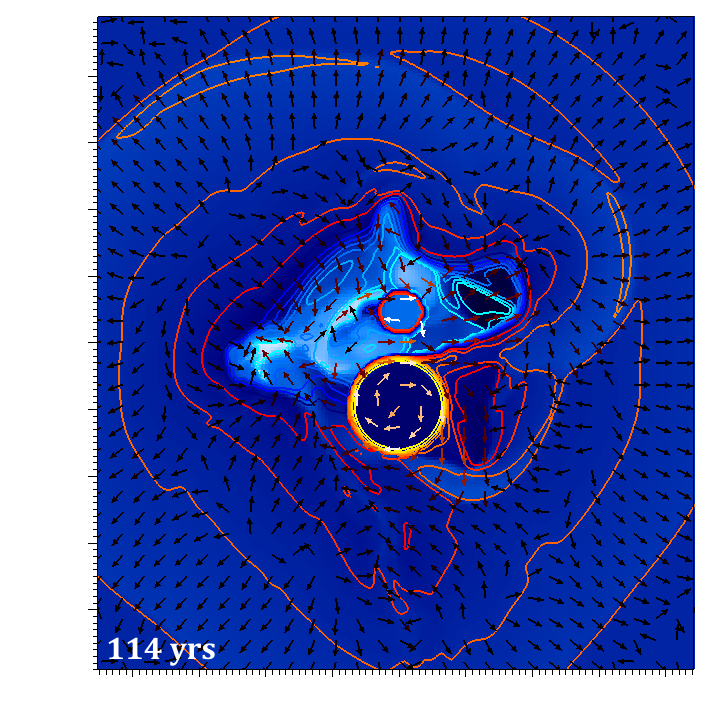}}
\subfloat{\includegraphics[width=0.20\linewidth]{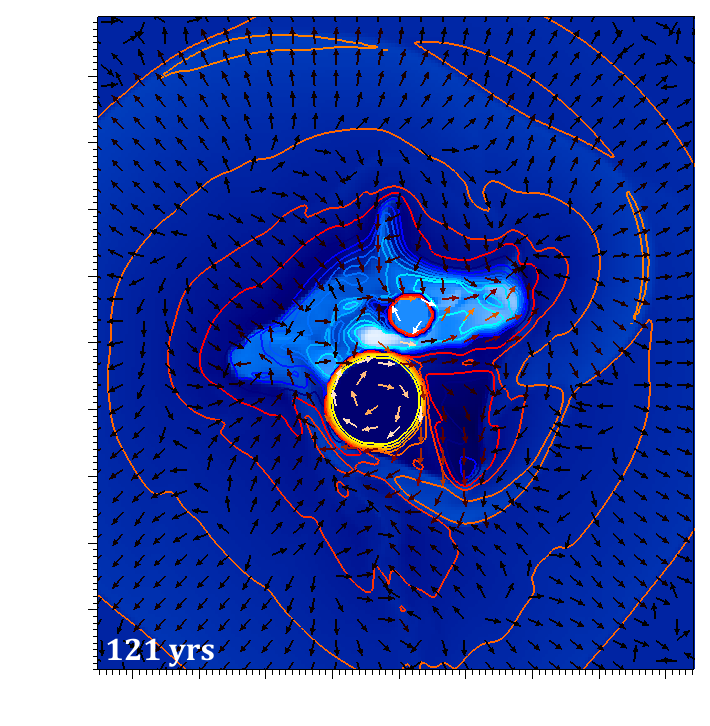}}
\subfloat{\includegraphics[width=0.20\linewidth]{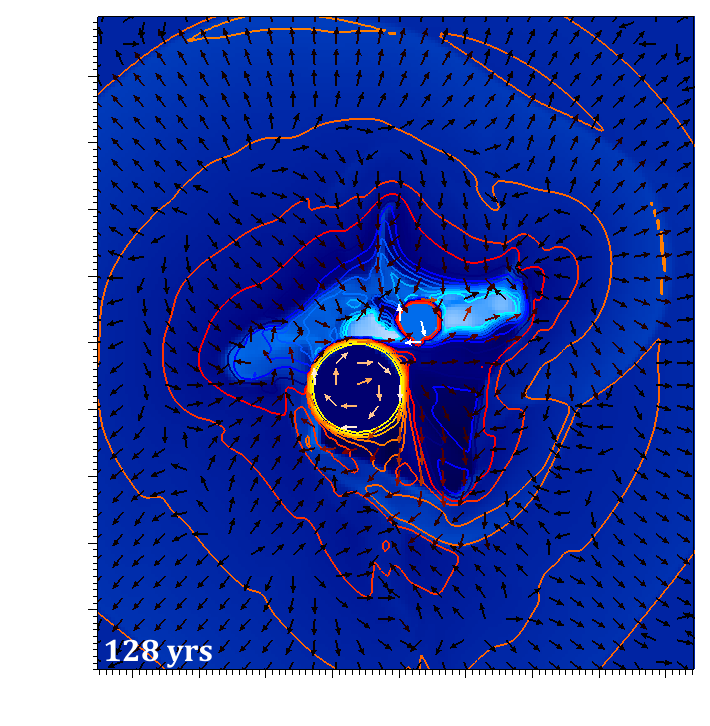}}
\subfloat{\includegraphics[width=0.20\linewidth]{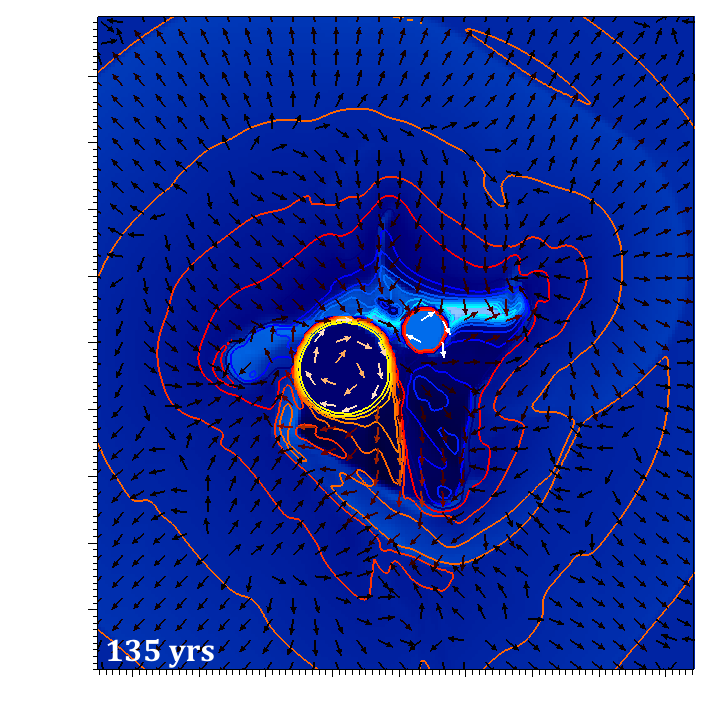}}\\[-2.5ex]
\subfloat{\includegraphics[width=0.20\linewidth]{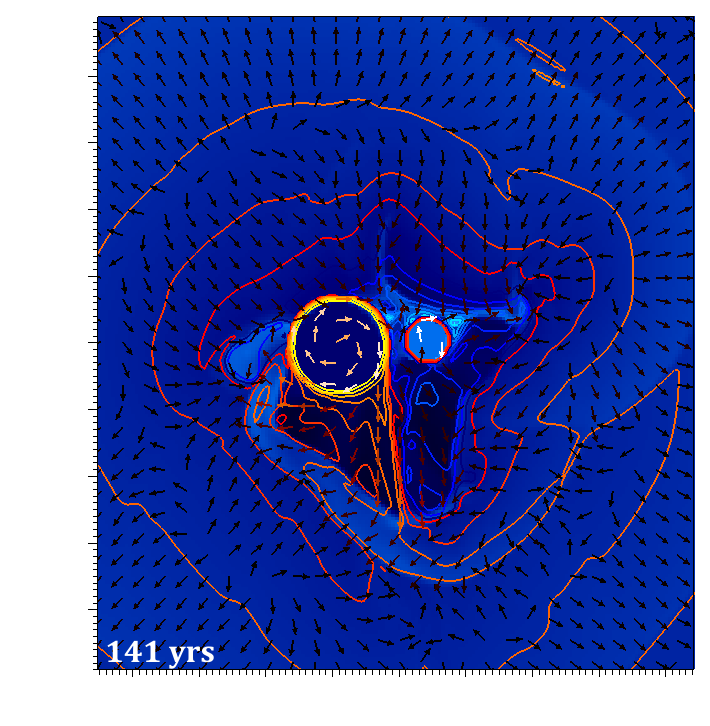}}
\subfloat{\includegraphics[width=0.20\linewidth]{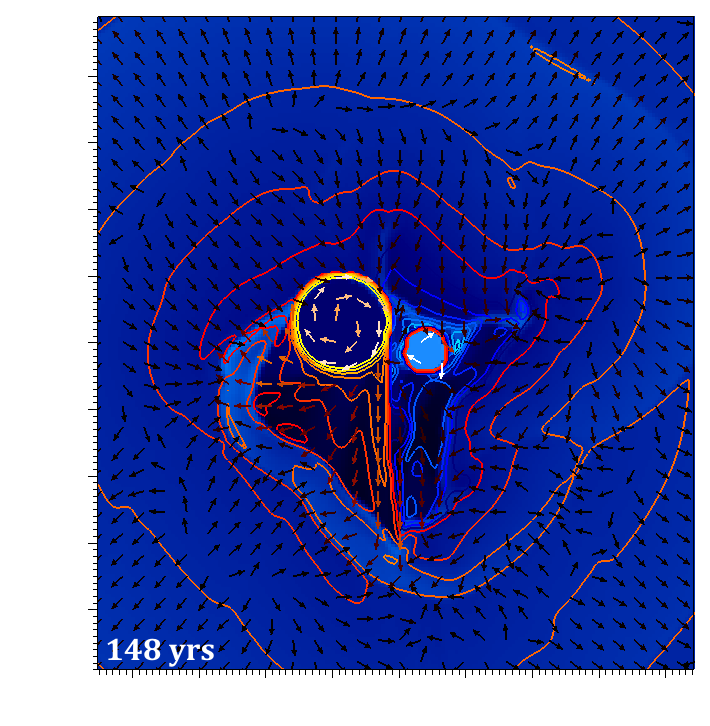}}
\subfloat{\includegraphics[width=0.20\linewidth]{./Images/delaval/a_6623_dln_23.png}}\\[-2.5ex]
\caption[18AU Atomic Primary, Molecular Secondary, Dynamic Interaction]{18AU Atomic Primary, Molecular Secondary, orbitally driven interaction; cross sections at x=3.2$\times$10$^{13}$ cm.  Background  (blue) shows temperature.  Velocity vectors are constant length, colour-scaled.  Contours show density.  Model number: s066.2.3}
\label{delaval}
\end{figure*}

The behaviour is nicely clarified in close-up in Figure \ref{delavalone} with the time sequence in  Figure \ref{delaval} which depicts  the two jet columns close to the inlets where these cavities form.  The cavity begins as the trailing wake of the molecular outflow, and as this outflow moves off, material is drawn from the outflow by the pressure differential.  The material rushes across this low-density region and meets the far `wall' where the pressure it exerts supports the cavity against collapse.  It gradually inflates radially outwards even as the continuing progress of the molecular outflow pushes open more wake in the ambient medium.

This situation does not persist indefinitely.  As the atomic jet swings around in its orbit it penetrates into the cavity.  The first effect of this is to disturb the flow and prevent further inflation of the cavity, though the flow separates in a portion of the cavity that carves off to form the starting point for a new cavity (see the '88 years' subfigure).  Then, some years later, a dramatic flaring event occurs as ionised material at 10$^4$ - 10$^5$\,K erupts from the vicinity of the atomic jet column and floods out into the low density cavity.  The process then begins again.

In addition to helping to understand the formation of cavities, Figure \ref{delaval} also demonstrates how the flow is  driven outwards into the ambient medium faster than the spiral wave; on passing through the forward shock the flow becomes disorganised as previously remarked on.

\subsection{Longitudinal Analysis of Co-orbital Models}
\label{longmethod}

Quantities of interest pertaining to the mainly atomic jet (including its neutral, ionised or entrained molecular hydrogen) were analysed as a function of distance along the (barycentric) x-axis of the problem domain, using IDLÂ® post-processing scripts developed for the purpose.  The fast-moving material could be isolated with a filter that selected for zones where V$_x$ was greater than 50 km~s$^{-1}$.  The exception was the analysis of ionised material which is simply selected material of that nature irrespective of velocity.

Generally, quantities examined were averaged over a slice of monozonal thickness for each of the 160 values of the x-coordinate, weighted by density or volume as appropriate, or in some cases the total quantity was determined.  The graphs presented here include plots of the atomic-molecular co-orbital model (s066.2.3) represented by solid lines, and of the atomic-only co-orbital model (s066.2.4) by dashed lines.  Given the density of data points, the single-valued nature of the functions and lack of meaningful error that may be attached, simple connected lines have been used rather than showing separate data points.

Given the fact that quantities are being averaged over the cross-sectional area of the jet, it is reasonable to extend this principle in the x-direction also, so that each data point represents the average over a three-dimensional region of the jet.  Therefore trend graphs smoothed in the x-direction are shown to the right of each plot.  Near-Gaussian smoothing is employed over a smoothing window equal to two jet inlet diameters.  This averages out the variations related to the velocity pulsations of the atomic jet.

 A dramatic change in the behaviour of the atomic jet when it  encounters the molecular outflow is apparent in nearly every case. These show the point of contact is at x $\approx$ 15\,AU from the inlet boundary.  In the smoothed graphs the effect shifts downstream, to around 20-25\,AU. 

\begin{figure*}
\includegraphics[width=0.55\linewidth, angle=90, bb=0 -74 504 610]{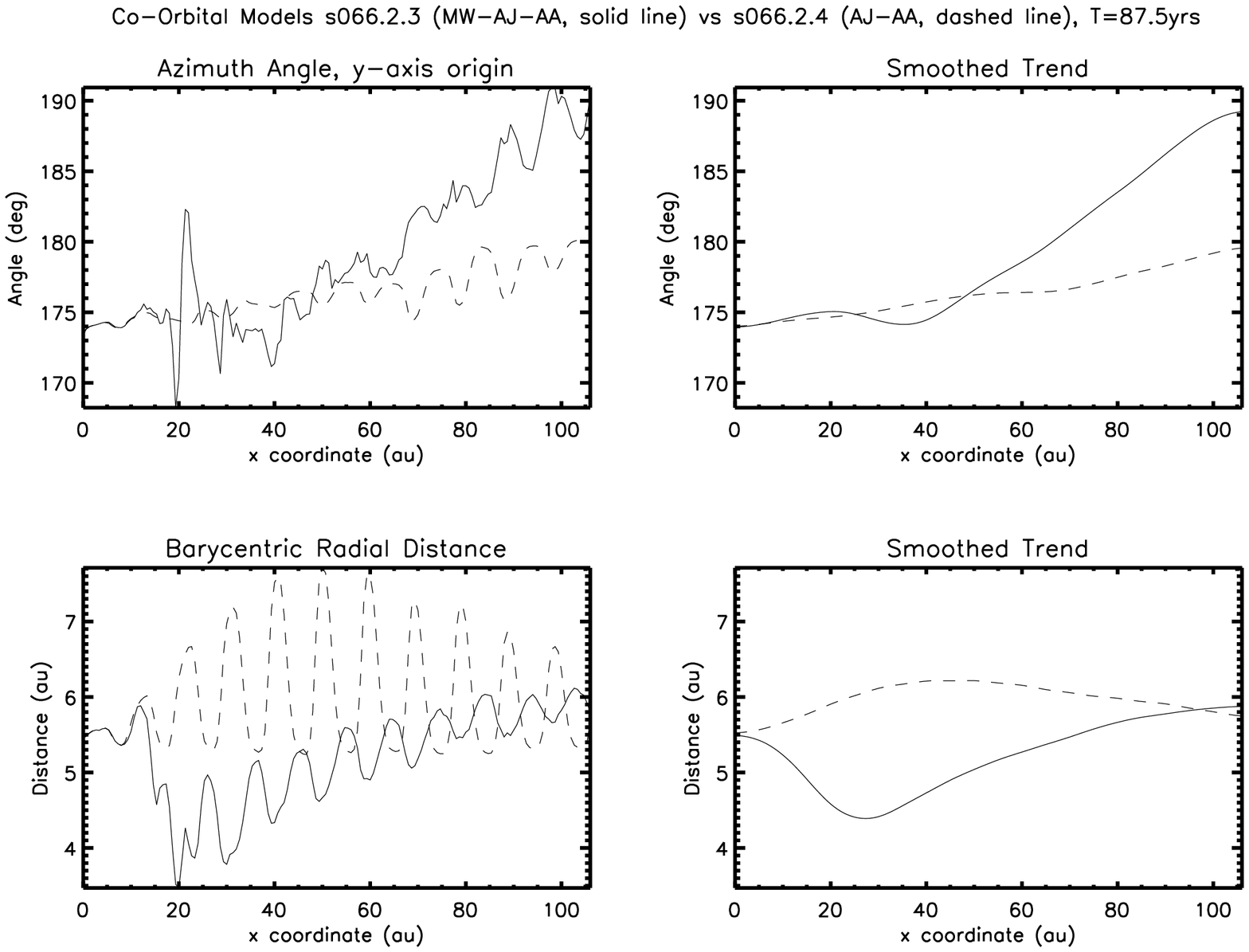}
\caption[Co-orbital Models: atomic jet azimuth angle and radial distance]{ Atomic jet azimuth angle and radial distance, along propagation axis, at 87.5 years for the co-orbital models.  Radial distance is in the y-z plane and directed outward with respect to the domain x-axis, which passes through the barycentre of the binary system.  Azimuth angle is in the y-z plane and directed in an anticlockwise sense about the domain x-axis.  Model numbers:  s066.2.3 (with molecular outflow) and s066.2.4 (without).}
\label{6623_jetbend_coords}
\end{figure*}

The atomic jet is deflected and twisted by the presence of the molecular jet. This is evident in
 the lower-right panel of Fig.\,\ref{6623_jetbend_coords}, where we see, perhaps surprisingly, an inward deflection towards the x-axis, before the jet is deflected outwards, as we might naturally expect.  The  explanation for this is that the atomic jet entrains molecular material at the point where it first encounters the dense molecular wind; thus sufficient x-momentum is imparted to some molecular material to push it into the $> 50$~km\,s$^{-1}$ velocity range we are selecting. This, thereby, suddenly weights the average position of the centre of mass closer to the radial origin.  After this initial inward deflection, we see an outwards deflection increasing monotonically, the radial limit of which may not be much greater than that of the undeflected atomic-only jet although a more elongated problem domain would be required to investigate this further.

However, Figure \ref{6623_jetbend_coords} shows that the azimuth deflection angle  attained by the end of the problem domain differs more conclusively from the unperturbed atomic jet (the azimuthal deflection of which is simply related to the jet's orbital motion). A difference of 9.5$^{\circ}\pm$0.1$^{\circ}$ is found..  This has implications for the winding ratio or pitch of the large-scale helical outflow beyond the short domain examined here.



\begin{figure*}
\includegraphics[width=0.55\linewidth, angle=90, bb=0 -74 504 610]{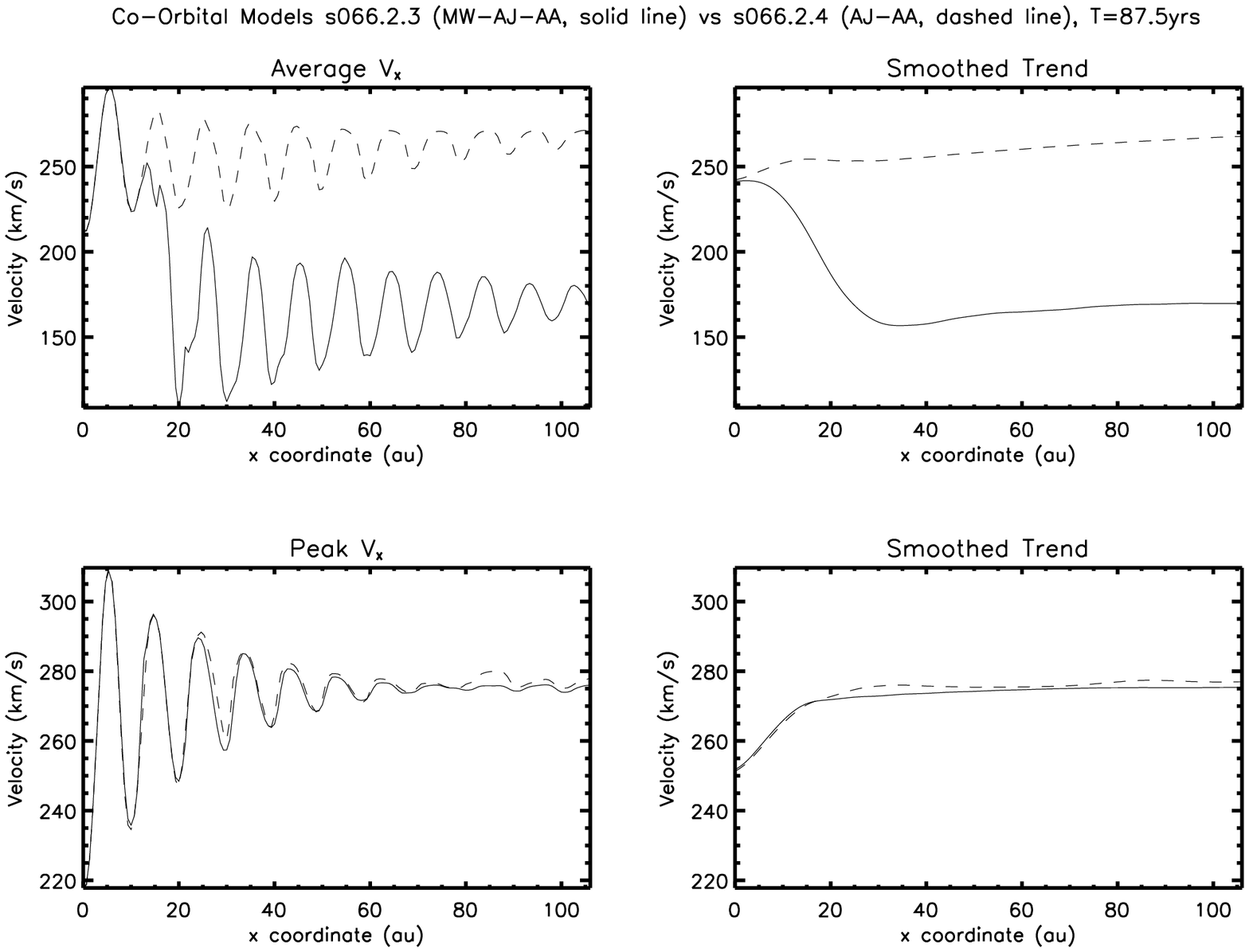}
\caption[Co-orbital Models: atomic jet average and peak velocity]{The average and peak velocity along the propagation axis of the atomic jet after 87.5 years.  Model numbers:  s066.2.3 (with molecular outflow, solid line) and s066.2.4 (without, dashed libe).}
\label{6623_jetbend_velocity}
\end{figure*}

The forward-directed velocity of the atomic jet is analysed in Figure \ref{6623_jetbend_velocity}.
 In the case of HH\,30, this lies almost in the sky plane for the first 400\,AU of propagation.  We see that there is a dramatic fall in the average velocity of the jet material from the point at which it encounters the molecular outflow. This is around 100 km\,s$^{-1}$ negative differential against a velocity of 250 - 270 km\,s$^{-1}$, remaining near-constant in the velocity smoothed trend until the jet exits the domain.  However, the peak velocity of material in the jet is virtually unchanged from the scenario where no molecular component is present.  This implies that there is a spine to the jet that may be bent but is not impeded.

\begin{figure*}
\includegraphics[width=0.55\linewidth, angle=90, bb=0 -74 504 610]{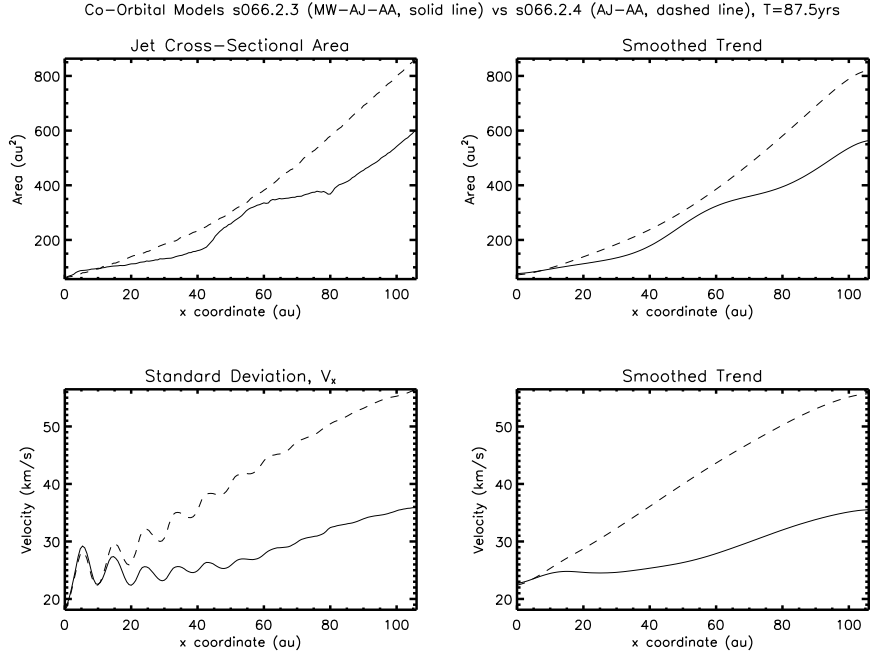}
\caption[Co-orbital Models: atomic jet x-sectional area and velocity dispersion]{Co-orbital Models: atomic jet x-sectional area and velocity standard deviation along propagation axis, T=87.5 years.  Model numbers:  s066.2.3 (with molecular outflow) and s066.2.4 (without).}
\label{6623_jetbend_misc}
\end{figure*}

\begin{figure*}
\centering
\subfloat{\includegraphics[width=0.55\linewidth, angle=90, bb=-30 -94 474 590]{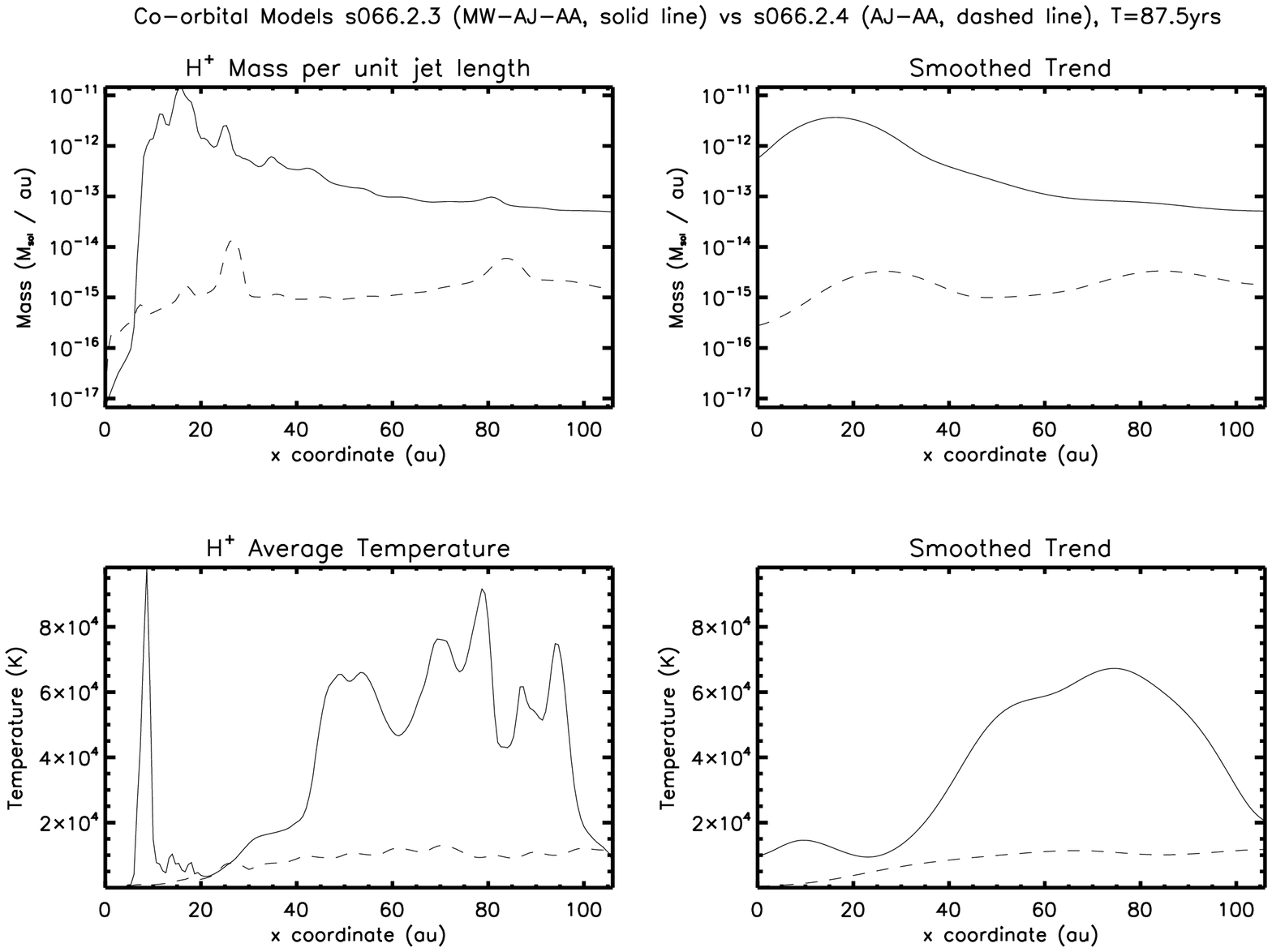}}
\hspace*{0.1\linewidth}\vspace{-2em}\caption[Co-orbital Models: thermal behaviour]{Ionisation and ion temperature along the jet propagation axis at 87.5 years.  Model numbers:  s066.2.3 (co-orbital with molecular outflow) and s066.2.4 (without).}
\label{6623_jetbend_thermal}
\end{figure*}

The width of the velocity distribution is displayed in Fig.\,\ref{6623_jetbend_misc}. This shows that the perturbed jet has been squeezed,  the jet cross-sectional area being only 65\% of the unperturbed jet.
 
\begin{figure*}
\hspace{2.2cm}\subfloat{\includegraphics[width=0.6\linewidth, angle=90, bb=-30 -94 474 590]{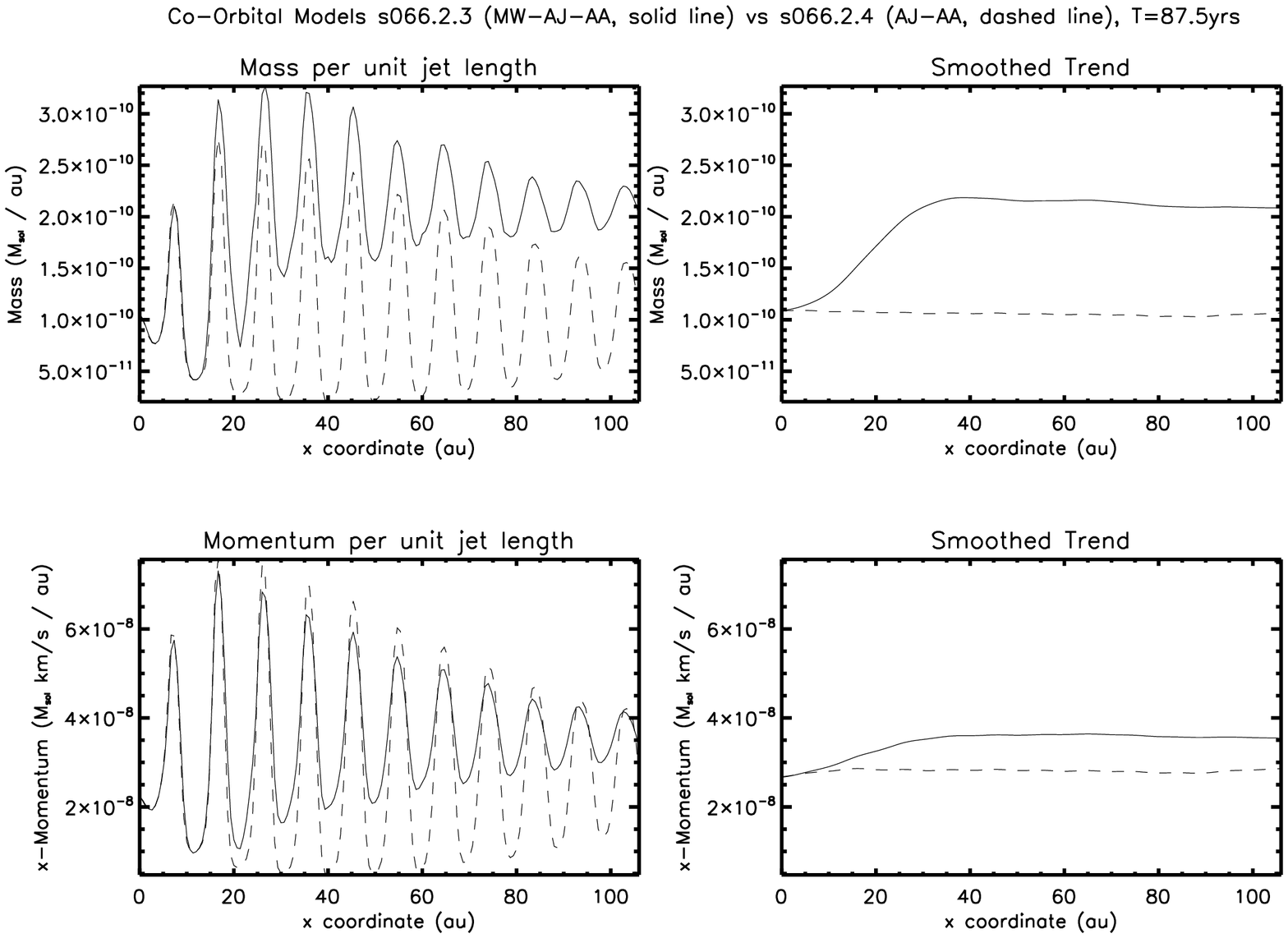}}\\
\hspace{0.2cm}
\subfloat{\includegraphics[width=0.6\linewidth, angle=90, bb=-30 -94 474 590]{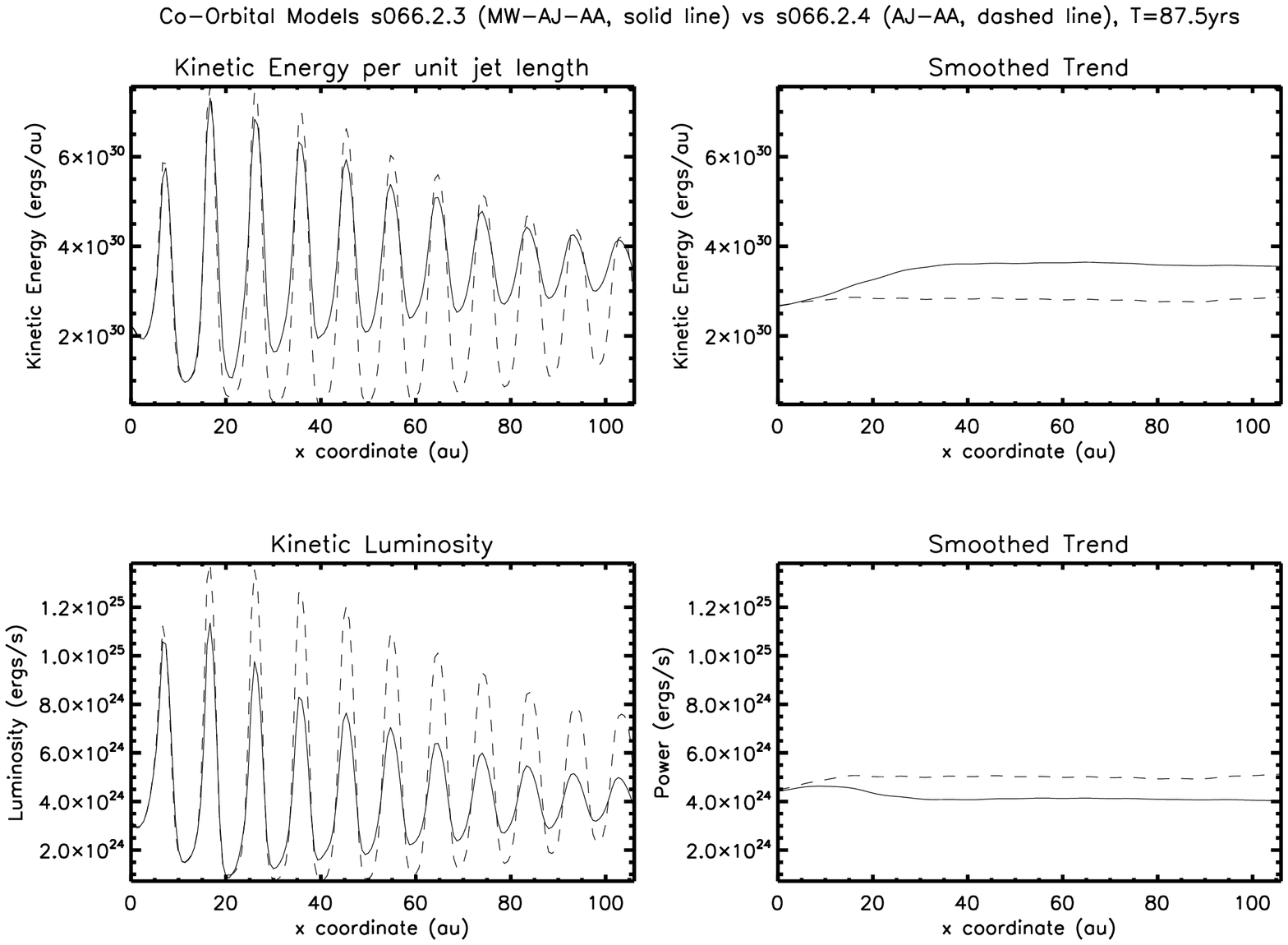}}\
\vspace{-2em}\caption[Co-orbital Models: mass and kinetic behaviour]{Atomic jet mass, momentum, kinetic energy and kinetic power along the propagation axis, at time 87.5 years.  Model numbers:  s066.2.3 (co-orbital with molecular outflow) and s066.2.4 (without).}
\label{6623_jetbend_momentum}
\end{figure*}

Figures  \ref{6623_jetbend_thermal} and  \ref{6623_jetbend_momentum}    show various other quantities, with self-evident differences between the perturbed and unperturbed jet.  
The ionisation caused by the interaction is extremely large.  As shown in  Fig.\,\ref{6623_jetbend_thermal},  within the first 20\,AU of propagation the perturbed jet produces an ionised mass 3-4 orders of magnitude greater than the unperturbed jet.  In the dual outflow co-orbital model it is the interaction between the outflows that produces the overwhelming majority of ions.  This increased ionisation persists out until the exit boundary by which time the difference has fallen to one order of magnitude.  

The mass per unit jet length shows a significant rise after the point of contact with the molecular outflow, approximately doubling its unperturbed value.  Since the momentum (in the same figure) has increased by $\sim$ 18\%,  we expect the velocity to reduce to $\sim$ 59\% of its value.  Referring to 
Fig.\,\ref{6623_jetbend_velocity}, an approximate calculation finds the velocity to have reduced by $\sim$ 58\%.  These calculations are approximate because the jet is not a closed system due to entrainment and mixing, and we do not expect exact conservation of momentum.
Nonetheless, we find it is still an almost-conserved quantity over the 100\,AU of our problem domain.

\section{Synthetic Imaging and maps}  
\label{synthetic}

Numerous atomic and molecular emission line properties can be calculated via post-processing of the physical variables. We have developed an IDL code , MULTISNTH, to generate images, position-velocity diagrams, channel mapes  and mass-velocity profiles These are  all useful as diagnostics for the specific model and full details will be presented in a future paper. 

\begin{figure*}
\subfloat{\includegraphics[width=0.48\linewidth]{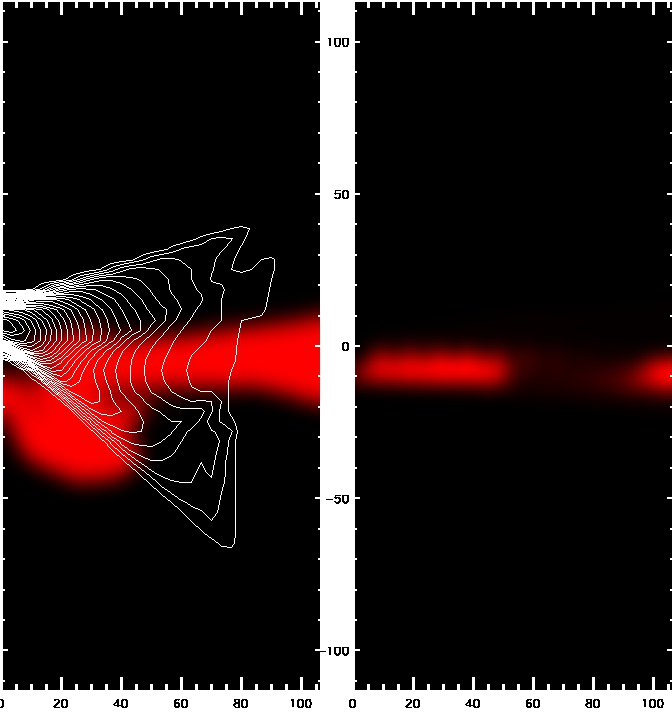}} \hfill 
\subfloat{\includegraphics[width=0.48\linewidth]{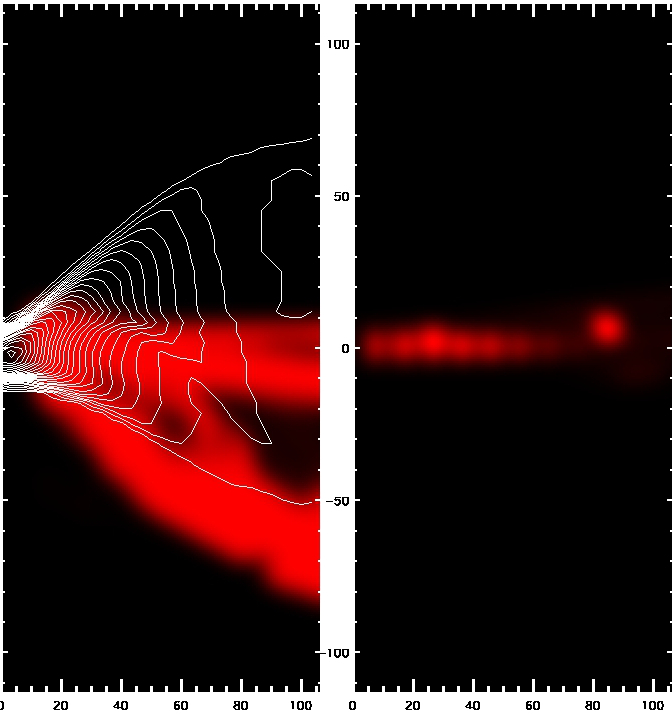}}\\
(a)  44 years \hspace{7.8cm} (b) 88 years \\
\subfloat{\includegraphics[width=0.48\linewidth]{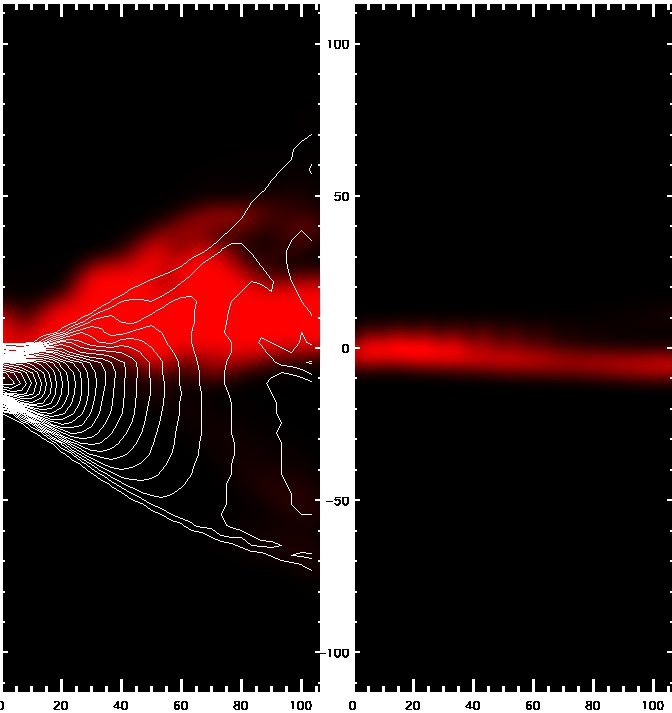}} \hfill 
\subfloat{\includegraphics[width=0.48\linewidth]{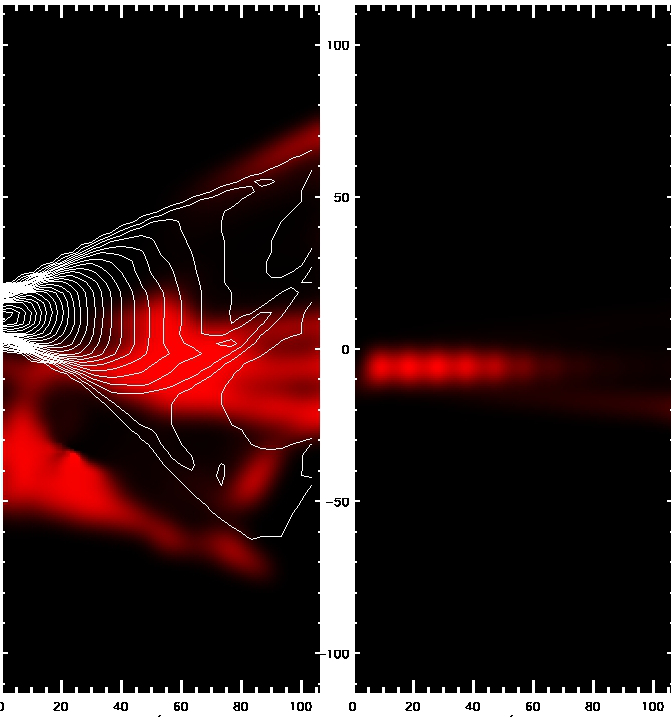}}\\
(a)  132 years \hspace{7.8cm} (b) 176 years \\
 \caption[Synthetic H-$\alpha$ Four Model Plot]
 {Synthetic H$\alpha$ images and CO contours  at the four indicated times from the co-orbital model with dual outflows (left sub-panels) and with just the atomic jet (right sub-panels)
 The optical emission is  smoothed with 14\,AU radius to match HST/WFPC2 pixel resolution of 0.1$\arcsec$ at 140\,pc.  Contours indicate molecular material (unsmoothed).}
\label{halpha}
\end{figure*}

Here, we present images of the H$\alpha$ 656\,nm emission superimposed on maps of the emission from the CO 2-1 rotational transition at 231\.GHz. This serves to highlight the interaction at the impact zone between the outflows. Figure\,\ref{halpha} displays the suitably smoothed H$\alpha$ images with CO contours.

The images demonstrate several defining features. Firstly, the undisturbed atomic jet is found to generate a prominent string of H$\alpha$ knots. The knots are more distinct  in some snapshots but tend to blend together at other times. The jet axis clearly deviates from the domain symmetry axis which may correspond to the underlying disc rotation axis.
Secondly, the appearance of the atomic jet is disrupted by the proximty of the molecular outflow in this model.  The heavy molecular outflow pushes hard on the atmin material, creating a warm bubble which is pushed out along the edge of the outflow. The molecular outflow impacts the jet at different angles, pushing the atomic material to different sides on the images. There is considerable delay between the two emissions on the scale of 100\,AU. As a result the H$\alpha$ and CO can be aligned or displaced.
The optical jet is considerably wider with individual knots not discernable.

The CO outflow maintains its character of a  limb brightened cone. At late times, there is some indication of entrainment into the jet which raises the possibility that molecular bullets could be catapulted out. However, it is not plausible to achieve high speeds before hydrodynamic instabilities disperse the entrained clumps.

\section{Conclusions}  
\label{conclusions}

We have discussed proposed interpretations of of interacting jets associated with young stellar objects. After comparing the parameters which have been suggested,
we then settled on two scenarios to take forward for detailed study. In this paper, we have restricted  the results to the co-orbital binary model in which an atomic jet and a molecular outflow are injected from discrete orbiting sources separated by a quite short distance.  The magnetic field is ignored. After some initial experiments, interesting field configurations within the jet presented computational inconsistencies.
The hydrodynamic simulation begins on scales of order 10\,AU whereas the field-driven launch  may be confined to within 0.1\,AU. Nevertheless, the field may remain crucial on the 10 -- 100\,AU scale.

The hot ($> 1,000$\,K) atomic jet was 10 $\times$ over-pressured with respect to ambient and modulated with a sinusoidal velocity pulse signal. The  period was 5.26 $\times$ 10$^6$ s.  The jet's minimum pulsed inlet velocity was 66\% of its maximum velocity of 326 km/s.  Further simulations were performed with different pulse characteristics. These are not included here but demonstrate that the pulse parameters do not alter the global flow dynamics.
An atomic ambient medium was assumed for the T\,Tauri close environment; ambient medium gradients and other inhomogeneities were eschewed as the modelled outflows would be run for enough simulation time to nurture their own domains.    

Our main working simulations necessarily covered a shorter span of jet propagation (107\,AU) than the early prototyping simulations due to the need to accommodate the wide-angle molecular flow within the limits of available computing resources, and the desire for better resolution.  The main atomic jet bow shock departed the problem domain well before the simulation time window in which results were calculated as we wished to examine the jet's steady-state behaviour.  

The co-orbital simulations are discussed via cross-section plots of physical variables to illustrate the dynamics.  In the absence of the molecular flow, the primary mode of ionisation is atomic jet material processed through internal working surfaces within the pulsed jet column.  When the co-orbital molecular flow is introduced, the main source of ionisation is the shock boundary between the atomic and molecular outflows.  The orbital dynamics of the two outflows produce some interesting structure in the surrounding medium, with low-density voids forming in the wake of the dense molecular outflow, which are then invaded and destroyed by the atomic outflow; this produces dramatic lateral flares of ionised material.  An analysis of the jet's longitudinal characteristics has shown various differences  between the perturbed and unperturbed atomic jet.

Synthetic images in atomic H$\alpha$ and molecular CO  lines are presented
 which demonstrate signatures specific to this model.
 In particular, the structure within the atomic jet is blurred and H$\alpha$ emission is strong from sections of the 
 walls of the CO cavity or from where the walls have been recently carved. 

A surrounding circumbinary  disc exists but is ignored here. This outer disc supplies the circumprimary disc which is ultimately reponsible for the fast atomic jet. In addition, it supplies the circum-secondary disc which feeds the heavier molecular outflow. 
We assume  the jet and outflow possess the same axial direction for the sake of these computations. However, some results can be generalised and allow us to speculate on how mis-aligned outflows may interact as their sources orbit. In certain geometries, the strongly  disturbed and ionised outflow may alternate on half the orbital period.  In the context of the co-orbital interpretation of HH\,30, a jet-counterjet  asymmetry would occur out to 400\,AU.
 Additional simulations with eccentric orbits were also performed in this study (see Table\,\ref{tab:simulationruns}) that show that the jet axial properties depend quite strongly on the chosen parameters.

Can the well-known HH\,30 asymmetries be produced here? Asymmetric changes in brightness in the outer circumbinary disc could have a number of origins
\citep{1996ApJ...473..437B,1999AAS...195.0202S}. One possibility raised is that of the passage of
clumps in the molecular wind \citep{2008MNRAS.387.1313T}. These intervening clumps  could be sufficiently dense to provide obscuration of stellar light through dust extinction. Here, on the scales simulated, the wind itself is optically thin. Figure\,\ref{6623_1300_xsect} does show enhancements in the outflow's molecular density where it gets compressed against the atomic jet. However, there is no clear evidence for clumping and, in addition,  compression in one dimension would not increase the dust column and hence the extinction.

A second known asymmetry is between the jet and counter-jet. Relevant to this work on the physical parameters, we can discuss the velocity variations, the counterjet displaying more variations in the speeds of the knots \citep{2012AJ....144...61E}.  
Here, we remark from Fig.\,\ref{6623_jetbend_velocity} that the passage of the atomic jet is hindered by the molecular outflow, reducing the axial flow speed considerably.
Moreover, as a result of the shock interaction, mixing and  dissociation, the jet luminosity falls but the momentum rises (Fig.\,\ref{6623_jetbend_momentum}).
 
 To conclude,  the wide molecular flow from the orbiting source significantly disturbs  the atomic jet, deflecting and twisting it  and disrupting the  dense knots. Orbiting regions of high ionisation are generated as the atomic jet rams through the molecular outflow. 
 In the next work, we  will place the atomic jet within the cone of the molecular outflow and  study the differences. 
 
These results provide a framework for the interpretation of upcoming sub-arcsecond  observations.    In particular, high spatial resolution long-slit spectroscopy of forbidden lines from space will be achievable for more than just the brightest microjets. Physical quantities can be deduced along the jet axis and through the knots as done for DG\,Tau by
\citet{2002ApJ...576..222B} and \citet{2014A&A...565A.110M} from Hubble Space Telescope data.  The new generation, led by
the James Webb Space Telescope and  the European Extremely Large Telescope, is capable of resolving structure on the scales of a few pixels as presented in the plots here.


\section*{Acknowledgements}  
\label{acks}

CL and MDS thank SEPnet and the University of Portsmouth for supplying infrastructure. 
SCOG acknowledges support from the Deutsche Forschungsgemeinschaft (DFG) via SFB 881 ``The Milky Way System'' (sub-projects B1, B2 and B8), 
and from the Heidelberg cluster of excellence EXC 2181 `STRUCTURES: A unifying approach to emergent phenomena in the physical world, mathematics, and complex 
data' funded by the German Excellence Strategy. 

\bibliography{thesis}

\appendix
\label{appendix}

\label{lastpage}

\end{document}